\documentclass[12pt]{article}
\usepackage{graphicx,amsmath,amssymb}
\usepackage{times}
\usepackage{pifont}
\newcommand{\amuh}{a_\mu^{\rm had}}
\newcommand{\dalh}{\Delta \alpha_{\rm had}}
\newcommand{\epm}{e^+e^-}
\newcommand{\gv}{\rm GeV }
\newcommand{\D}{\rm d}
\newcommand{\epo}{\;.}
\newcommand{\power}[1]{\times 10^{#1}}

\newcommand{\nn}{\nonumber}

\newcommand{\nc}{\newcommand}
\newcommand{\gapprox}{\raisebox{-.2ex}{$\stackrel{\textstyle>}{\raisebox{-.6ex}[0ex][0ex]{$\sim$}}$}}

\newcommand{\mytexttilde}{{\raise.17ex\hbox{$\scriptstyle\mathtt{\sim}$}}}
\nc{\pipi}{\pi^+\pi^-}
\nc{\be}{\begin{equation}}
\nc{\ee}{\end{equation}}
\nc{\ra}{\rightarrow}
\nc{\gam}{\gamma \gamma}
\nc{\br}{\eta \ra \pi^0 \gam}
\nc{\bb}{\bibitem}
\nc{\omg}{\omega}
\nc{\g}{\gamma}
\nc{\ba}{\begin{eqnarray}}
\nc{\ea}{\end{eqnarray}}
\nc{\mv}{{\rm MeV }}
\nc{\I}{{\rm i} }
\renewcommand{\D}{{\rm d} }
\nc{\E}{{\rm e} }
\nc{\cF}{{\cal F} }
\nc{\cM}{{\cal M} }
\nc{\cR}{{\cal R} }
\nc{\tM}{\tilde{\cM}}

\newcommand{\mbo}[1]{$#1$ }
\newcommand{\crn}{\nonumber \\}
\begin{document}
\begin{titlepage}
\vbox{~~~ \\[-12mm]
                                  \null \hfill \small{LPNHE/2016--01}\\
                                   \null \hfill \small{DESY 16--081}\\
\title{A BHLS model based moment analysis of muon $g-2$,
 \\ and its use for lattice QCD evaluations of $\amuh$.
   }
\author{
M.~Benayoun$^a$, P.~David$^{a,b}$, L.~DelBuono$^a$, F.~Jegerlehner$^{c,d}$ \\
\small{$^a$ LPNHE des Universit\'es Paris VI et Paris VII, IN2P3/CNRS, F--75252 Paris, France }\\
\small{$^b$ LIED, Universit\'e Paris-Diderot/CNRS UMR 8236, F--75013 PARIS, France  } \\
\small{$^c$ Humboldt--Universit\"at zu Berlin, Institut f\"ur Physik, Newtonstrasse 15, D--12489 Berlin,
Germany }\\
\small{$^d$ Deutsches  Elektronen--Synchrotron (DESY), Platanenallee 6, D--15738 Zeuthen, Germany}
}
\date{\today}
\maketitle
\begin{abstract}
We present an up-to-date analysis of muon $g-2$ evaluations in terms
of Mellin-Barnes moments as they might be useful for lattice QCD
calculations of $a_\mu$.  The moments up to 4th order are evaluated
directly in terms of $\epm$--annihilation data and improved within the
Hidden Local Symmetry (HLS) Model, supplied with appropriate symmetry
breaking mechanisms. The model provides a reliable Effective
Lagrangian (BHLS) estimate of the two-body channels plus the
$\pi\pi\pi$ channel up to 1.05~GeV, just including the $\phi$
resonance. The HLS piece accounts for 80\% of the contribution to
$a_\mu$. The missing pieces are evaluated in the standard way directly
in terms of the data. We find that the moment expansion converges well
in terms of a few moments. The two types of moments which show up in
the Mellin-Barnes representation are calculated in terms of hadronic
cross--section data in the timelike region and in terms of the
hadronic vacuum polarization (HVP) function in the spacelike region
which is accessible to lattice QCD (LQCD). In the Euclidean the first
type of moments are the usual Taylor coefficients of the HVP and we show that 
the second type of moments may be obtained as integrals over the appropriately
Taylor truncated HVP function.  Specific results for the isovector part 
of $\amuh$ are determined by means of HLS model predictions in close relation to
$\tau$--decay spectra.
\end{abstract}
}
\end{titlepage}

\section{Introduction to the moments expansion approach}
\label{sec:introduction}
In the lattice QCD (LQCD) approach of calculating $\amuh$,
extrapolation methods have been developed (see e.g. contributions
to~\cite{Benayoun:2014tra}) to overcome difficulties to reach the
physical point in the space of extrapolations. The low $Q^2$ behavior
of the Euclidean electromagnetic current correlators on a lattice,
which exhibits a discrete momentum spectrum, poses a particular
challenge (see e.g.~\cite{Boyle:2011hu,Aubin:2013daa} and references
below). Actually, $Q^2=0$ is not directly accessible, because of the
finite volume, which represents an infrared (IR) cutoff. The analysis of moments of the
subtracted (i.e. renormalized) photon vacuum polarization function
$\Pi(Q^2)=e^2\,\hat{\Pi}(Q^2)$ ($e$ the positron charge) was
particularly advocated in variants in Refs.~\cite{deDivitiis:2012vs}
and \cite{Aubin:2012me}. Recent lattice
calculations~\cite{Feng:2013xsa,Francis:2014dta,DellaMorte:2014rta,Malak:2015sla,Bali:2015msa}
have been utilizing moment analysis techniques for a more precise
evaluation of $\amuh$. The leading moment is given by the slope of the
Adler function~\cite{deRafaelEJLN94} as follows from the
representation:
\be
\amuh=\frac{\alpha^2\,m_\mu^2}{6\pi^2}\, \int\limits_0^1 \D x\:x\:(2-x)\:
\left(D(Q^2(x))/Q^2(x)\right)
\label{ADI}
\ee
with $Q^2(x)\equiv \frac{x^2}{1-x}m_\mu^2$ the spacelike square
momentum transfer, and $D(Q^2)$ the Adler function, defined as a
derivative of the shift of the fine structure constant $\Delta
\alpha_{\mathrm{had}}(s)\equiv -4\pi\alpha\,\hat{\Pi}(s)$:
\be
D(-s)=-(12\pi^2)\,s\,\frac{\D \hat{\Pi}(s)}{\D s}
=\frac{3\pi}{\alpha} s\frac{d}{\D s}\Delta \alpha_{\mathrm{had}}(s) \epo
\label{DD}
\ee
The Adler function is represented by\footnote{We somewhat sloppy
write $s_0=4 m_\pi^2$ for the lower integration limit, which is the
threshold for the dominating $\pi^+\pi^-$ channel.  However, the true
threshold for contributions to $R(s)$ is $s_0=m_{\pi^0}^2$ as $e^+e^- \to \pi^0 \gamma$
is the process exhibiting the lowest threshold. Most lattice QCD
simulations for simplicity are done for the isovector piece, where $4
m_\pi^2$ is the correct threshold.}:
\ba
D(Q^2)=Q^2\:\left(\int_{4 m_{\pi}^2}^{\infty}\frac{
R(s)}{(s+Q^2)^2}\D s\:\right)
\label{DI}
\ea
in terms of $R(s)$, which can be evaluated in
terms of experimental $\epm$ data as well as, to a large part, in terms of our HLS
model prediction. The Adler-function $D(Q^2)$ is bounded asymptotically by perturbative
QCD (pQCD): $D(Q^2)\to N_c\,\sum_f Q_f^2$, with $Q_f$ the quark charges and
$N_c=3$ the color factor, up to perturbative corrections, which
asymptotically vanish because of asymptotic freedom which implies
$\alpha_s(Q^2) \to 0$ as $Q^2 \to \infty$
(see~\cite{EJKV98}). Obviously, then $D(Q^2)/Q^2$ is
a positive monotonically decreasing function bounded by:
\ba
\frac{D(Q^2)}{Q^2}=\int_{4 m_{\pi}^2}^{\infty}\frac{
R(s)}{(s+Q^2)^2}\D s\: < D'(0)\equiv \int_{4 m_{\pi}^2}^{\infty}\frac{
R(s)}{s^2}\D s\:=\left.\frac{D(Q^2)}{Q^2}\right|_{Q^2=0}\;,
\ea
the slope of the vacuum polarization function at zero momentum
square. The finite slope guarantees the convergence of the integral
(\ref{ADI}) at the lower limit. For our analysis it is important to
know how the integrand of (\ref{ADI}) looks like, in order to know
where the important contributions show up.
\begin{figure}[h]
\centering
\includegraphics[width=0.45\textwidth]{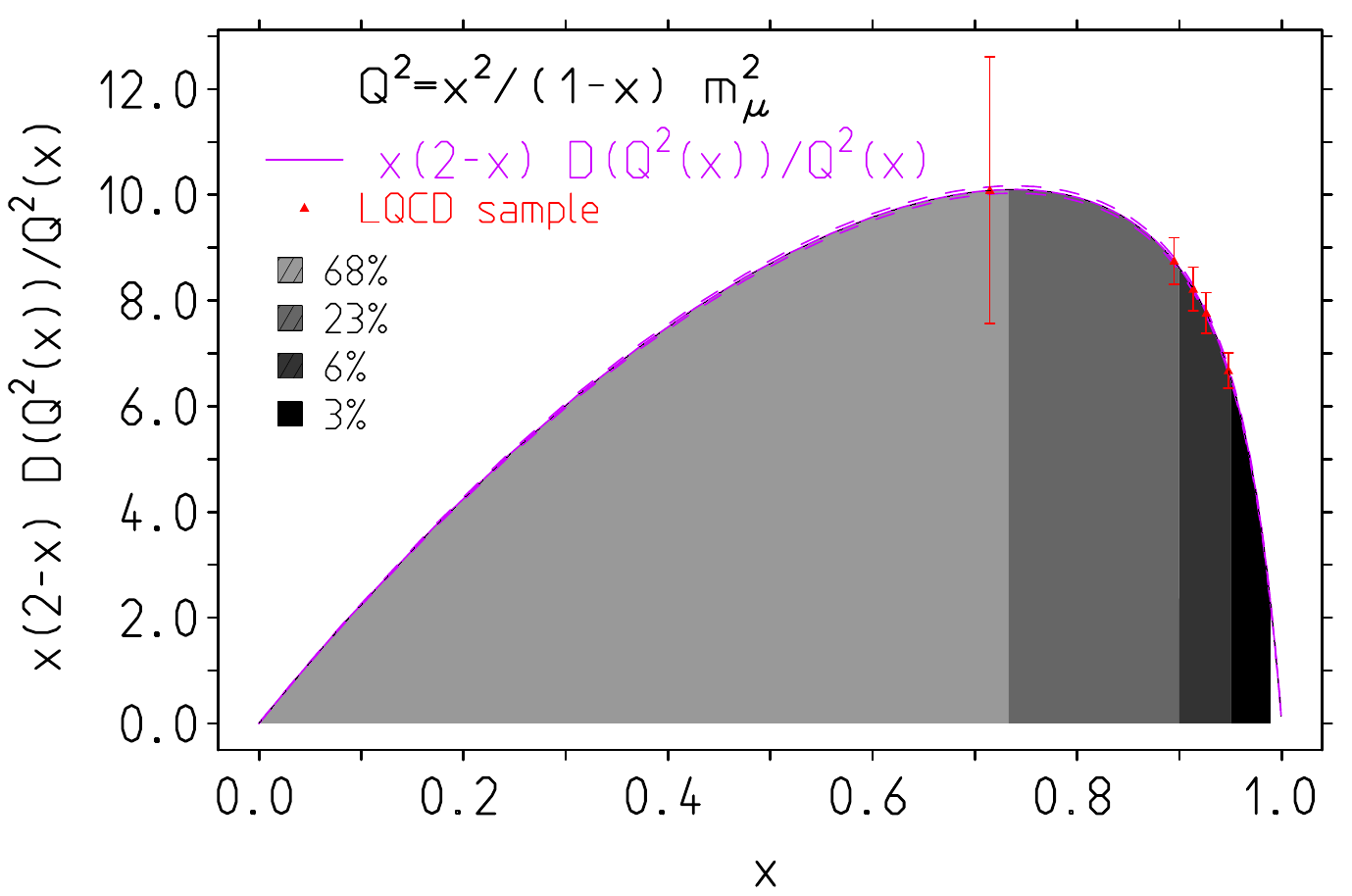}
\includegraphics[width=0.45\textwidth]{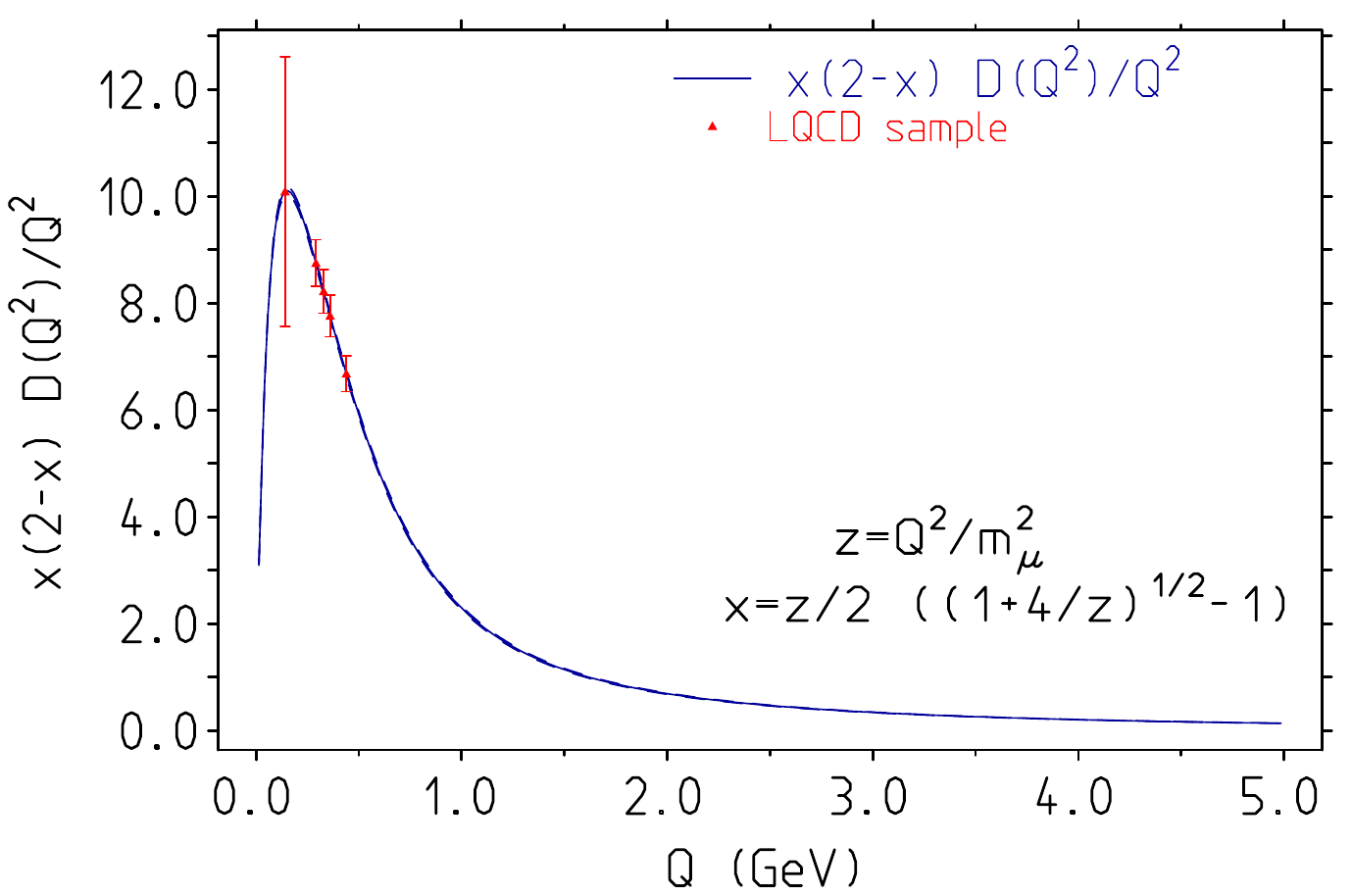}
\caption{The integrand of the Adler function representation (\ref{ADI}) as a function of $x$ and as a
function of the energy scale $Q$. The right--hand panel shows that the integrand is sharply
peaked as a function of $Q$ at a rather low
scale ($\sim 150~\mv$). Adler function data  come from ~\cite{Adlerfunction}.
The dashed lines mark the error band from the experimental data. ``LQCD sample''
shows points of $Q_{\rm min}$ from Ref.~\cite{Aubin:2015rzx} presently
achievable in lattice QCD simulations (shown are pseudo-data lying on the curve, we assumed a 25\% uncertainty
for the lowest point and
a 5\% uncertainty for the higher ones). In the left panel we also display
the contributions to $\amuh$ from
regions between $Q_i=0.00,\,0.15,\,0.30,\,0.45$ and $1.0~\gv$ in
percent. The tail above 1~GeV contributes slightly less than 0.2\%.}
\label{fig:ADIkernels} 
\end{figure}
Figure~\ref{fig:ADIkernels} shows a pronounced peak at a surprisingly
low scale of about $Q \approx 150~\mv$. This shows that the dominant $\rho$
contribution appears to be shifted towards lower scales
in the Euclidean region.

For the slope, using (\ref{DD}), we may write: 
\be
D'(0)=-\frac{3\pi}{\alpha} \frac{\D}{\D s}\Delta
\alpha_{\mathrm{had}}(s)|_{s=-Q^2,Q^2\to 0}=12\,\pi^2\, \frac{\D}{\D s}\,\hat{\Pi}(s)|_{s=-Q^2,Q^2\to 0}
\ee
directly as the slope of the photon self-energy function
$\Pi(Q^2)\equiv 4\pi\alpha \hat{\Pi}(s)$. An evaluation in terms of
data yields
\be
D'(0)\simeq 10.20(7)~\gv^{-2}\epo
\ee
The Adler function slope $D'(0)$ has been estimated in lattice QCD
in~\cite{Francis:2013qna}. The LQCD result $D'(0)=5.8(5)~\gv^{-2}$ has
been compared with $D'(0)=9.81(30)~\gv^{-2}$, a result obtained using
a phenomenological toy-model representation~\cite{Bernecker:2011gh} of
the isovector spectral function. As another example we mention the
result $D'(0)=12\pi^2 \sum_{u,d,s,c}
Q_f^2\times\Pi_1=10.67(17)[9.95(17)]~\gv^{-2}$ we get with
$\Pi_1=0.0811(12)[0.0756(13)]$ obtained in
Ref.~\cite{Chakraborty:2016mwy} for set 8[10] of Table~II, the 
closest to the physical
point. The lattice results usually include the isovector
part only, which is simpler but difficult enough, and are often
missing some higher energy contributions above 1~GeV.

Note that (\ref{ADI}) is equivalent to the standard formula:
\be
\amuh=\left(\frac{\alpha\,m_\mu}{3\pi}\right)^2\,
\int\limits_{s_0}^\infty\frac{\D s}{s^2}\:\hat{K}(s)\:R(s)
\label{lohadalt}
\ee
in which $\hat{K}(s)$ is a bounded monotonically increasing function,
with $\hat{K}(4m_\pi^2)\simeq 0.63$ going to 1 as $s\to
\infty$. Setting $\hat{K}(s)=1$ we obtain a true upper bound (see also~\cite{deRafaelEJLN94}):
\be
\amuh < \left(\frac{\alpha m_\mu}{3\pi}\right)^2 \:D'(0)< 784(6)\power{-10}\epo
\label{VPbound}
\ee
The result is way too large as the dominant low energy part of $R(s)$
is obviously overweighted. A lower bound is obtained by setting
$\hat{K}(s)=\hat{K}(4m_\pi^2)\approx 0.63$, which implies $\amuh > 494
(4)\power{-10}$, again a very rough bound only, but a true bound. These
bounds can be much improved by a systematic low energy expansion of
the kernel function in (\ref{ADI})\footnote{A low energy expansion of
the kernel of (\ref{lohadalt}) is by far not straightforward as we
have to deal with the $2m_\mu$ threshold of $\hat{K}(s)\,.$}, as
advocated recently in Ref.~\cite{deRafael:2014gxa}, specifically as a
tool to get more precise results from the Euclidean
lattice data. It provides a novel approach for evaluating $\amuh$ in
terms of moments, which goes beyond a simple Taylor expansion of
$\Pi(Q^2)$, where the latter, as such, can be integrated only in the
range of validity of the expansion. The starting point here is the
Mellin-Barnes representation:
\be
\amuh=\left(\frac{\alpha}{\pi}\right)\,\frac{1}{2\pi \I}\: \int\limits_{c-\I \infty}^{c+\I
\infty}\,\D s\: \cF(s)\,\cM(s)
\label{MBrep}
\ee
with the exact analytic kernel
\ba
\cF(s)&=&-\Gamma(3-2s)\,\Gamma(-3+s)\,\Gamma(1+s)\,,
\ea
in terms of Euler Gamma functions $\Gamma(s)$.  The function $\cM(s)$
is the Mellin transform of the hadronic spectral function:
\be
\cM(s)=\frac{\alpha}{3\pi}\int\limits_{4m_\pi^2}^{\infty}\:\frac{\D t}{t}\,R(t)\,
\left(\frac{m_\mu^2}{t}\right)^{1-s}\,,
\label{MellinHVP}
\ee
and allows  to perform a moment expansion by weighting $R(t)$ with
powers of $m_\mu^2/t$, as it appears in (\ref{MellinHVP}).
Remember that $\Gamma(s)$ is a
meromorphic function of $s$ with simple poles at $s=-n$
($n=0,1,2,\cdots$) and residues $(-1)^n/n!$.  $\Gamma(x)$ is real
positive for real positive values of $x$. The pole structure, which resides on
the closed negative real axis, then follows from repeated applications
of $\Gamma(s)=\Gamma(s+1)/s$, until $s+1$ is positive.

The low momentum expansion proposed in~\cite{deRafael:2014gxa} is
derived by calculating the residues of the poles of the above
representation (\ref{MBrep})$\,$: $\cF(s)$ exhibits simple poles at
$s=0,-1,-2,
\cdots$ and double poles at $s=-1,-2, \cdots\,$:
\ba
\cF(s)&\simeq& \frac{1}{3}\,\frac{1}{s}-\frac{1}{(s+1)^2}
+\frac{25}{12}\,\frac{1}{s+1}-\frac{6}{(s+2)^2}
+\frac{97}{10}\frac{1}{s+2}\crn &&-\frac{28}{(s+3)^2}+
\frac{208}{5}\,\frac{1}{s+3}
-\frac{120}{(s+4)^{2}}+\frac{3608}{21}\,\frac{1}{s+4}
+ \cdots
\ea

The simple poles yield values:
\be
\cM(-n)=\frac{\alpha}{3\pi}\int\limits_{4m_\pi^2}^{\infty}\:\frac{\D s}{s}\,R(s)\,
\left(\frac{m_\mu^2}{s}\right)^{1+n}\epo
\label{Mint}
\ee
The double poles of $\cF(s)$ also require the first
derivative of the Mellin transform:
\be
\tilde{\cM}(-n)=\frac{\alpha}{3\pi}\int\limits_{4m_\pi^2}^{\infty}\:\frac{\D s}{s}\,R(s)\,
\,\ln(\frac{m_\mu^2}{s})
\left(\frac{m_\mu^2}{s}\right)^{1+n}
= -\frac{\D}{\D s}\left.\cM(s)\right|_{s=-n} \epo
\label{tildeMdire}
\ee
In terms of the moments, the successive approximations then read:
\ba
\label{tildeMint}
\begin{array}{cccl}
\amuh{(0)}&=&
&\left(\frac{\alpha}{\pi}\right)\,\left[\frac{1}{3}\cM(0)\right] \\
\amuh{(1)}&=&\amuh{(0)}+&\left(\frac{\alpha}{\pi}\right)\,\left[\frac{25}{12}\cM(-1)
+\tilde{\cM}(-1)\right]\\
\amuh{(2)}&=&\amuh{(1)}+&\left(\frac{\alpha}{\pi}\right)\,\left[\frac{97}{10}\cM(-2)
+6\tilde{\cM}(-2)\right]\\
\amuh{(3)}&=&\amuh{(2)}+&\left(\frac{\alpha}{\pi}\right)\,\left[\frac{208}{5}\cM(-3)
+28\tilde{\cM}(-3)\right]\\
\amuh{(4)}&=&\amuh{(3)}+&\left(\frac{\alpha}{\pi}\right)\,\left[\frac{3608}{21}\cM(-4)
+120\tilde{\cM}(-4)\right]\,,
\end{array}
\ea
Note that
$\cM(0)$ corresponds to the Adler function slope $D'(0)$ in
(\ref{VPbound}) as $\cM(0)=\frac{\alpha}{3\pi}\,m_\mu^2\,D'(0)$.

The analysis presented in the following includes $\epm$ annihilation
data from Novosibirsk (NSK) ~\cite{CMD203,CMD206,SND06}, Frascati
(KLOE)~\cite{KLOE08,KLOE10,KLOE12}, Stanford (BaBar)~\cite{BABARpipi}
and Beijing (BES\-III)~\cite{BESIII}, $\tau$-decay data from ALEPH,
OPAL, CLEO and Belle~\cite{ALEPH,AlephCorr,OPAL,CLEO,Belle}. Other data
on exclusive channels recently collected, and published up to the end
of 2014, include the $e^+e^-\to 3(\pi^+\pi^-)$ data from CMD--3
\cite{Akhmetshin:2013xc}, the $e^+e^- \to \omega\pi^0
\to \pi^0\pi^0\gamma$ from SND \cite{Achasov:2013btb} and several data
sets collected by BaBar in the ISR mode\footnote{Including the $p
\bar{p}$,~$K^+K^-$,~$K_LK_S,\:K_LK_S\pi^+\pi^-$,~$K_SK_S\pi^+\pi^-,K_S
K_S K^+K^-$ final
states.}~\cite{Lees:2013ebn,Lees:2013gzt,Lees:2014xsh,Davier:2015bka}.

In the following we present results for the moments directly in terms
of $\epm$ annihilation data as well as global fit improved BHLS
estimates for moments up to the 4th order, which allow us to get good
estimates for the full results. The main results, the evaluations of
the moments $\cM(-n)$ and $\tilde{\cM}(-n)$ for $n=0,1,2,3,4$ and the
corresponding results for $\amuh(n)$, are presented in Sect.~2. While
the moments $\cM(-n)$ are directly calculable by the Euclidean methods
of lattice QCD, the moments $\tilde{\cM}(-n)$ are only indirectly accessible 
in the Euclidean world. Therefore, in Sect.~3 we perform a
related calculation in terms of other types of moments, denoted by
$\Sigma(-n; s_0)$, which are directly accessible to lattice QCD
calculations and allow one to estimate $\tM(-n)$ as linear
combinations of $\Sigma(-n; s_0)$'s and of higher order $\cM(-n)$'s. As
indicated, the auxiliary moments $\Sigma(-n; s_0)$'s depend on an
infrared cutoff $s_0$ which should cancel in the linear combination
which corresponds to $\tM (-n)$, which by definition is independent of
$s_0$. In Sect.~4 we have a closer look on the method studied in
Sect.~3, which follows the line proposed in
Ref.~\cite{deRafael:2014gxa}. A more careful consideration reveals
that also the log suppressed moments $\tM(-n)$ can be obtained
directly from Euclidean momentum space by a limiting procedure $s_0
\to 0$ of an integral over a truncated HVP function. Sect.~5 
is devoted to the extraction of the I=1 part of the HVP function and
its contribution to the various moments.  It also provides more
details concerning the role of $\tau$-decay spectral data and the
isospin breaking effects in the HLS model. Isospin breaking effects,
and particularly vector meson mixing effects together and with the
photon, have to be included in order to properly relate the isovector
$\tau$ data to $\epm$ annihilation data. This provides insight into
the model dependence of the BHLS evaluations. Conclusions are
presented in Sect.~6.  For recent summaries of $\amuh$ evaluations we
refer to~\cite{Benayoun:2015hzf,Zhang:2015yfi,Jegerlehner:2015stw}.

\section{A BHLS model based moment analysis of $\amuh$ }
\label{sec:BHLSmoments}
Previous studies~\cite{Benayoun:2015gxa,Benayoun:2012wc} have shown
that the Hidden Local Symmetry (HLS) Model, supplied with appropriate
symmetry breaking mechanisms, provides an Effective Lagrangian (BHLS)
which encompasses a large number of processes within a unified
framework. A global fit procedure has been derived herefrom which
allows for a simultaneous description of the $e^+ e^-$ annihilation
into 6 final states -- $\pi^+\pi^-$, $\pi^0\gamma$, $\eta \gamma$,
$\pi^+\pi^-\pi^0$, $K^+K^-$, $K_L K_S$ -- and includes the dipion
spectrum in the $\tau$ decay and some more light meson decay partial
widths. The contribution to the muon anomalous magnetic moment
$a_\mu^{\rm th}$ of these annihilation channels over the range of
validity of the HLS model (up to 1.05~GeV) is found much improved
compared to the standard approach of integrating the measured $\epm$
spectra directly~\cite{Benayoun:2015gxa,Benayoun:2012wc}.  The key
point is that besides implementing the vector meson dominance model
(VDM) in accord with the chiral structure of QCD, the model allows to
treat the mixing of $\rho$, $\omega$, and $\phi$ among them and with
the $photon$ in a coherent way as a consequence of the vector meson
self-energy effects, which at the same time models the vector meson
widths and the related decays, and in particular models the
relationship  between $e^+e^-$ annihilations and
 the charged $\tau$ channel. In contrast to the
standard approach of integrating the $\epm$ data, the BHLS approach
incorporates the $\tau$ spectral data as a key ingredient. This
provides a welcome reduction of the leading hadronic uncertainty in the
lowest order (LO) hadronic vacuum polarization (HVP) to $a_\mu$. Here
we apply our approach to the moments analysis of $\amuh$.

\begin{figure}[h]
\centering
\includegraphics[height=7cm]{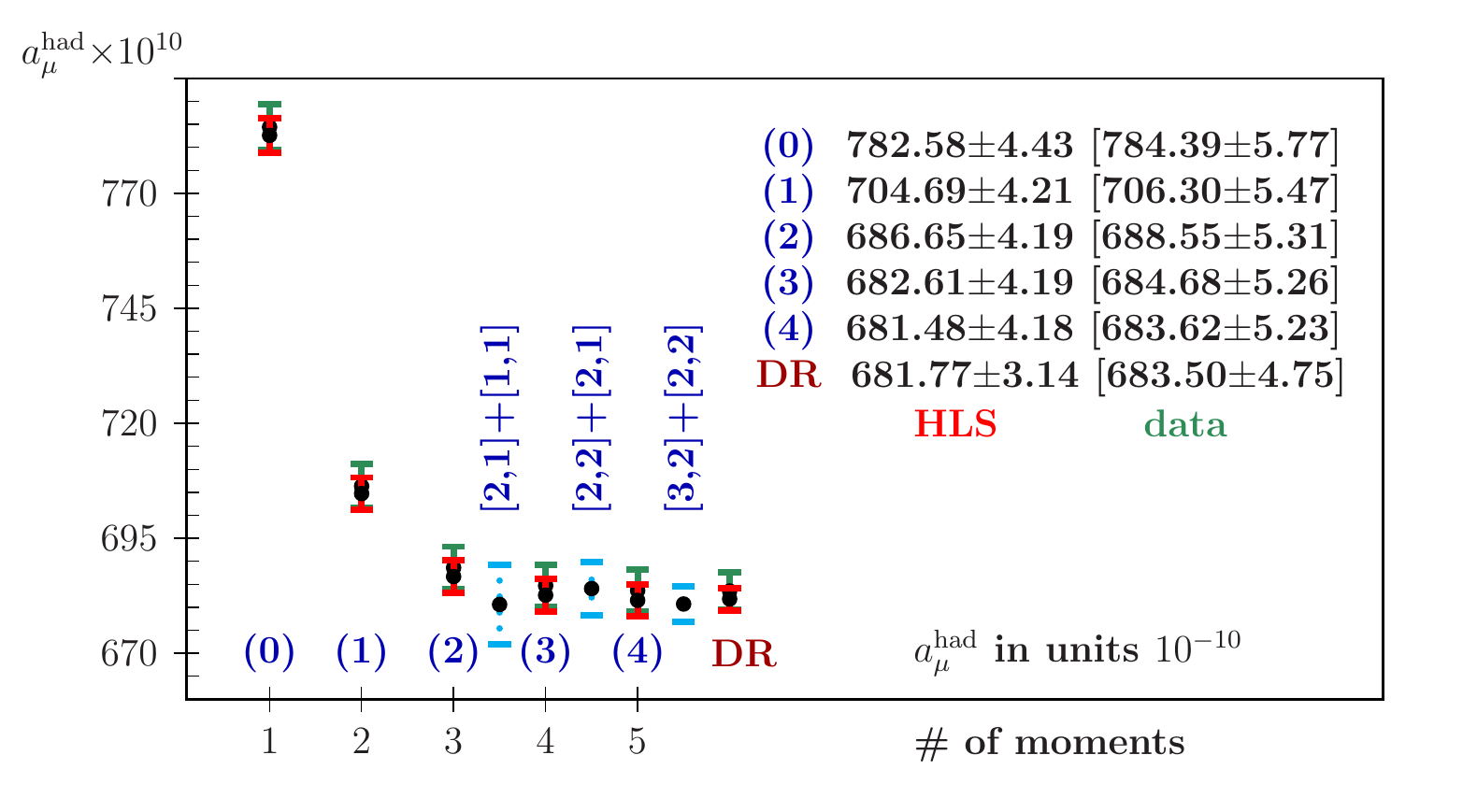}
\caption{Starting with a crudely overestimated approximation, the
successive higher moments converge rapidly. Shown are results from
Table~\ref{tab:momres} for
``data direct'' and ``HLS + remainder''. DR marks the result obtained
with the dispersion relation (\ref{lohadalt}). Shown are also the
Taylor-Pad\'e estimates of Table~\ref{tab:TaylorPades} based on 3, 4 and 5
Taylor coefficients.}
\label{fig:gm2moments} 
\end{figure}

\begin{center}
\begin{table}[h]
\caption{Moments of the $\amuh$-expansion in units $10^{-5}$. Here
$\cM(-n)$ and $\tilde{\cM}(-n)$ are evaluated via Eqs. (\ref{Mint})
and (\ref{tildeMdire}) in terms of $R(s)$ as provided by
$e^+e^-$ annihilation data and/or predictions of the BHLS model
Lagrangian. The ``data HLS channels'' denote the channels separated
from the $R(s)$ ``data direct'', which are predicted by means of the HLS
effective Lagrangian after determining its parameters by a global
fit. The prediction ``HLS model'' is then combined with the remainder
represented by the difference of the first two columns in column 4 as
``HLS + remainder''. Note the remarkable gain in accuracy when
replacing the ``data HLS channels'' by the ``HLS model'' prediction. The
improvement gets the better the higher the moment is, since higher moments
are more and more dominated by the low energy tail covered by the HLS model.}
\label{tab:mom}
{\scriptsize
\begin{tabular}{lr@{.}lr@{.}l|r@{.}lr@{.}l|r@{.}lr@{.}l|r@{.}lr@{.}l}
\noalign{\smallskip}\hline\noalign{\smallskip}
moment &
\multicolumn{4}{c}{data direct} &
\multicolumn{4}{c}{data HLS channels} &
\multicolumn{4}{c}{HLS model} &
\multicolumn{4}{c}{HLS + remainder}  \\
\noalign{\smallskip}\hline\noalign{\smallskip}
$\cM(0)  $
&10&1307&$\!\!\!\!\!\pm$0&0745
& 8&6275&$\!\!\!\!\!\pm$0&0495
& 8&6041 &$\!\!\!\!\!\pm$0&0130
&10&1073 &$\!\!\!\!\!\pm$0&0572	  \\
$\cM(-1) $
& 0&23507&$\!\!\!\!\!\pm$0&00185
& 0&22944&$\!\!\!\!\!\pm$0&00184
& 0&23197&$\!\!\!\!\!\pm$0&00031
& 0&23760&$\!\!\!\!\!\pm$0&00038   \\
$\cM(-2) $
& 0&008702&$\!\!\!\!\!\pm$0&000115
& 0&008669&$\!\!\!\!\!\pm$0&000115
& 0&008974&$\!\!\!\!\!\pm$0&000011
& 0&009007&$\!\!\!\!\!\pm$0&000011 \\
$\cM(-3) $
& 0&0004852&$\!\!\!\!\!\pm$0&0000093
& 0&0004850&$\!\!\!\!\!\pm$0&0000093
& 0&0005147&$\!\!\!\!\!\pm$0&00000064
& 0&0005149&$\!\!\!\!\!\pm$0&00000064\\
$\cM(-4) $
& 0&00003676&$\!\!\!\!\!\pm$0&00000083
& 0&00003676&$\!\!\!\!\!\pm$0&00000083
& 0&00003956&$\!\!\!\!\!\pm$0&00000005
& 0&00003956&$\!\!\!\!\!\pm$0&00000005 \\
\hline\noalign{\smallskip}
$\tilde{\cM}(-1)$
&-0&82592&$\!\!\!\!\!\pm$0&00516
&-0&79611&$\!\!\!\!\!\pm$0&00501
&-0&80054&$\!\!\!\!\!\pm$0&00113
&-0&83035&$\!\!\!\!\!\pm$0&00168 \\
$\tilde{\cM}(-2)$
&-0&026808&$\!\!\!\!\!\pm$0&000294
&-0&026644&$\!\!\!\!\!\pm$0&000294
&-0&027338&$\!\!\!\!\!\pm$0&000035
&-0&027503&$\!\!\!\!\!\pm$0&000035\\
$\tilde{\cM}(-3)$
&-0&0013160&$\!\!\!\!\!\pm$0&0000228
&-0&0013149&$\!\!\!\!\!\pm$0&0000228
&-0&0013847&$\!\!\!\!\!\pm$0&0000017
&-0&0013858&$\!\!\!\!\!\pm$0&0000017   \\
$\tilde{\cM}(-4)$
&-0&00009064&$\!\!\!\!\!\pm$0&00000199
&-0&00009063&$\!\!\!\!\!\pm$0&00000199
&-0&00009725&$\!\!\!\!\!\pm$0&00000012
&-0&00009726&$\!\!\!\!\!\pm$0&00000012 \\
\noalign{\smallskip}\hline
\end{tabular}
}
\end{table}
\end{center}
An up-to-date evaluation of the moments, based on a $R(s)$
compilation of $e^+e^-$ annihilation data together with the results
using the BHLS predictions is presented in Table~\ref{tab:mom}. The
improvement obtained by modeling the channels encompassed by the BHLS
model is what we observe going from ``data HLS channels'' to ``HLS
model'', which then is supplemented by the part not covered by the
effective Lagrangian in its range of validity to obtain the best
evaluation for ``HLS + remainder''. The evaluation of $\amuh$ in terms
of these moments follows in Table~\ref{tab:momres}. The bottom entries
are the results obtained with the exact kernel, as presented in the
previous section. The errors of the moments are 100\% correlated
although weighted differently for different energy regions. Since the
signed errors are added linearly with weight unity, errors apparently
get somewhat underestimated\footnote{Adding errors quadratically would
be simply wrong here. $\amuh$ is dominated by the $\pipi$ channel and
there by the systematic error. The dominance of the $\pipi$ channel
gets even more pronounced the higher the moment. As all moments are
linear in $R(s)$, if $R$ goes up the $\cM(-n)$'s go up and all
$\tM(-n)$'s go down. The different weighting for different $n$ does
not make them independent. Rather the $\cM$'s are close to 100\%
correlated while the $\tM$'s s are 100\% anti--correlated relative to
the $\cM$'s. Our error estimate is to be considered as an educated
guess.}.  One observes a nice convergence provided all contributions
are collected appropriately. One should keep in mind that the
dominating $\rho$ resonance accounts for about 75\% and the
predictable HLS channels account for about 80\% of $\amuh$.  Thus,
obviously, the non-HLS contribution including data at higher energies
(beyond our 1.05~GeV breakpoint) is important in getting the complete
results. Figure~\ref{fig:gm2moments} illustrates the fast convergence
of the first few moments, despite the fact that the lowest
order moment is quite far off. Therefore, in cases where the relevant
moments are available rather than $R(s)$, e.g. in lattice QCD, the
Mellin-Barnes moments approach, suggested in
Ref.~\cite{deRafael:2014gxa}, provides a reliable method for the
evaluation of $\amuh$.
\begin{table}
\centering
\caption{The LO-HVP contribution in terms of moments in units $10^{-10}$}
\label{tab:momres}
{\footnotesize
\begin{tabular}{lc|c|c|c}
\noalign{\smallskip}\hline\noalign{\smallskip}
\phantom{moment} &
{data direct} &
{data HLS channels} &
{HLS model} &
{HLS + remainder}  \\
\noalign{\smallskip}\hline\noalign{\smallskip}
 $\amuh(0)$&  784.39$\pm$    5.77&  668.00$\pm$    3.83&  666.19$\pm$    1.01&  782.58$\pm$  4.43\\
 $\amuh(1)$&  706.30$\pm$    5.47&  594.11$\pm$    3.56&  592.50$\pm$    0.89&  704.69$\pm$  4.21\\
 $\amuh(2)$&  688.55$\pm$    5.31&  576.51$\pm$    3.41&  574.62$\pm$    0.87&  686.65$\pm$  4.19\\
 $\amuh(3)$&  684.68$\pm$    5.26&  572.65$\pm$    3.35&  570.58$\pm$    0.87&  682.61$\pm$  4.19\\
 $\amuh(4)$&  683.62$\pm$    5.23&  571.59$\pm$    3.33&  569.45$\pm$    0.86&  681.48$\pm$  4.18\\
\noalign{\smallskip}\hline\noalign{\smallskip}
  $\amuh$  &  683.50$\pm$    4.75&  570.68$\pm$    3.67&  568.95$\pm$    0.89&  681.77$\pm$  3.14\\
\noalign{\smallskip}\hline
\end{tabular}
}
\end{table}

It is worthwhile to add a comment about the HLS model estimates of the
moments. While the direct data evaluation is based simply on weighted
averages of data sets which then are integrated using the trapezoidal
rule, the HLS model results are obtained by the Monte Carlo method.
The BHLS global fit using {\tt Minuit} provides: i/ the vector
$\vec{x}$ of the central values of the fit parameters ii/ the error
covariance matrix $V$. These are treated as the parameters of a
multidimensional Gaussian distribution $G(x,V)$, of which one performs
$N$ (a few hundreds or thousands) samplings.  For each sampling one
calculates the function $R(s)$. So, we have numerically $N$ estimates
of the function $R(s)$: $R(s,i)\, ,\,\, i=1,...N$ all defined in steps
of 0.5~MeV from threshold to 1.05~GeV.  One then can estimate any moment
$$Q= \int K_Q(s)\, R(s)\, ds\,,$$ with the appropriate kernel
$K_Q(s)$, to obtain sequences $$Q(i) = \int K_Q(s)\, R(s,i)\, ds,\,
i=1...N\,.$$ Then, the $Q(i)$ sequence can be histogrammed and fitted by a
Gaussian from which one derives the central value and the standard
deviation (this is done easily within {\tt Paw}). 

For what concerns
the data direct approach, we use chiral perturbation theory to
parametrize a fit of the low energy tail of the available data
(including timelike as well as spacelike data up to 
400~MeV)~\cite{Eidelman:1995ny}. Integrals are then performed adopting 
318~MeV as a ``chiral cut'', below which the fit is used in place of the
data compilation. Since the higher moments are completely dominated by
the HLS channels and the high energy tail above 1.05~GeV also gets
negligible, the higher moments directly reflect the difference of the
HLS model versus the standard data direct approach.

\section{Moments accessible in lattice QCD calculations}
\label{sec:LQCDmoments}
\begin{figure}[h]
\centering
\includegraphics[width=0.45\textwidth]{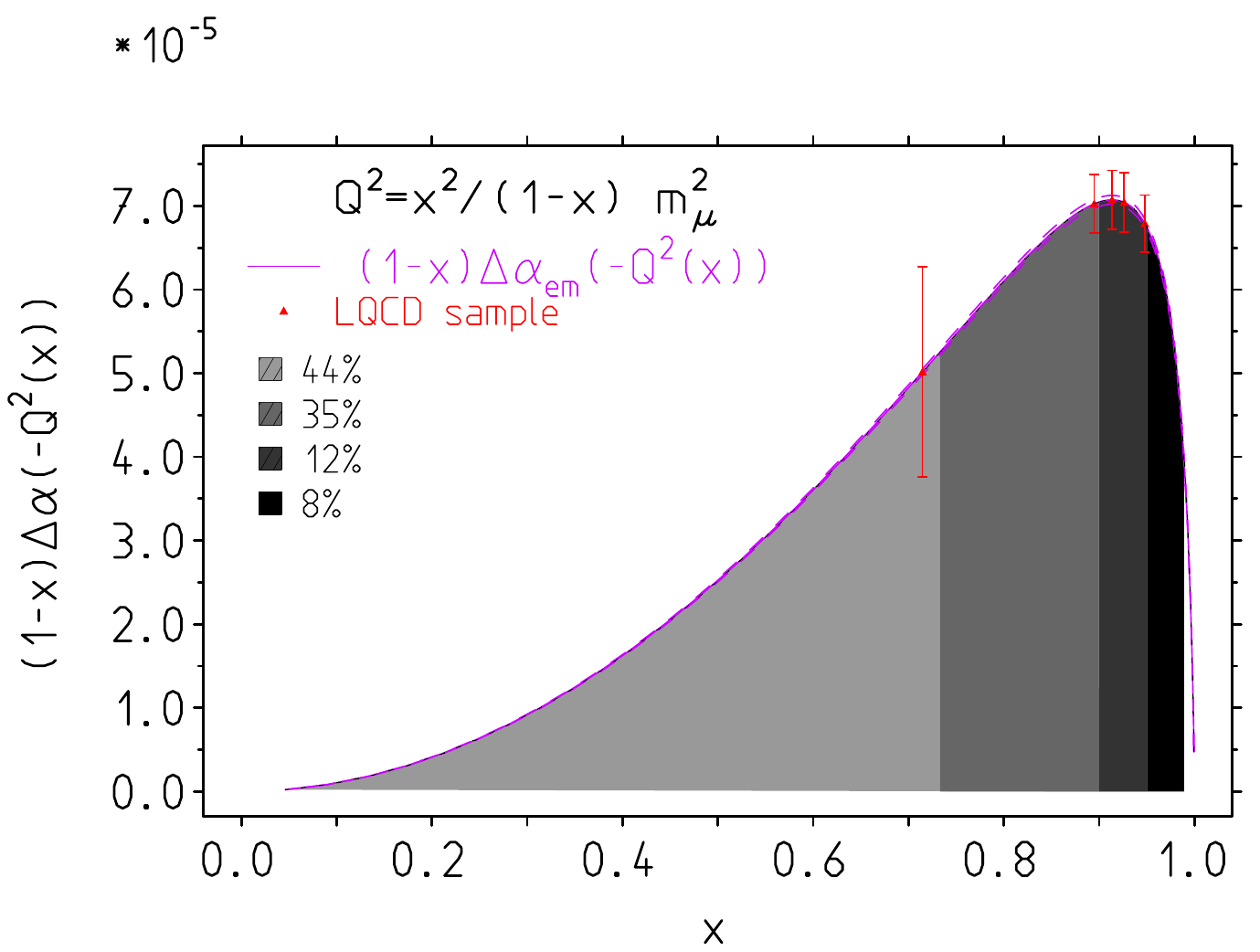}
\includegraphics[width=0.45\textwidth]{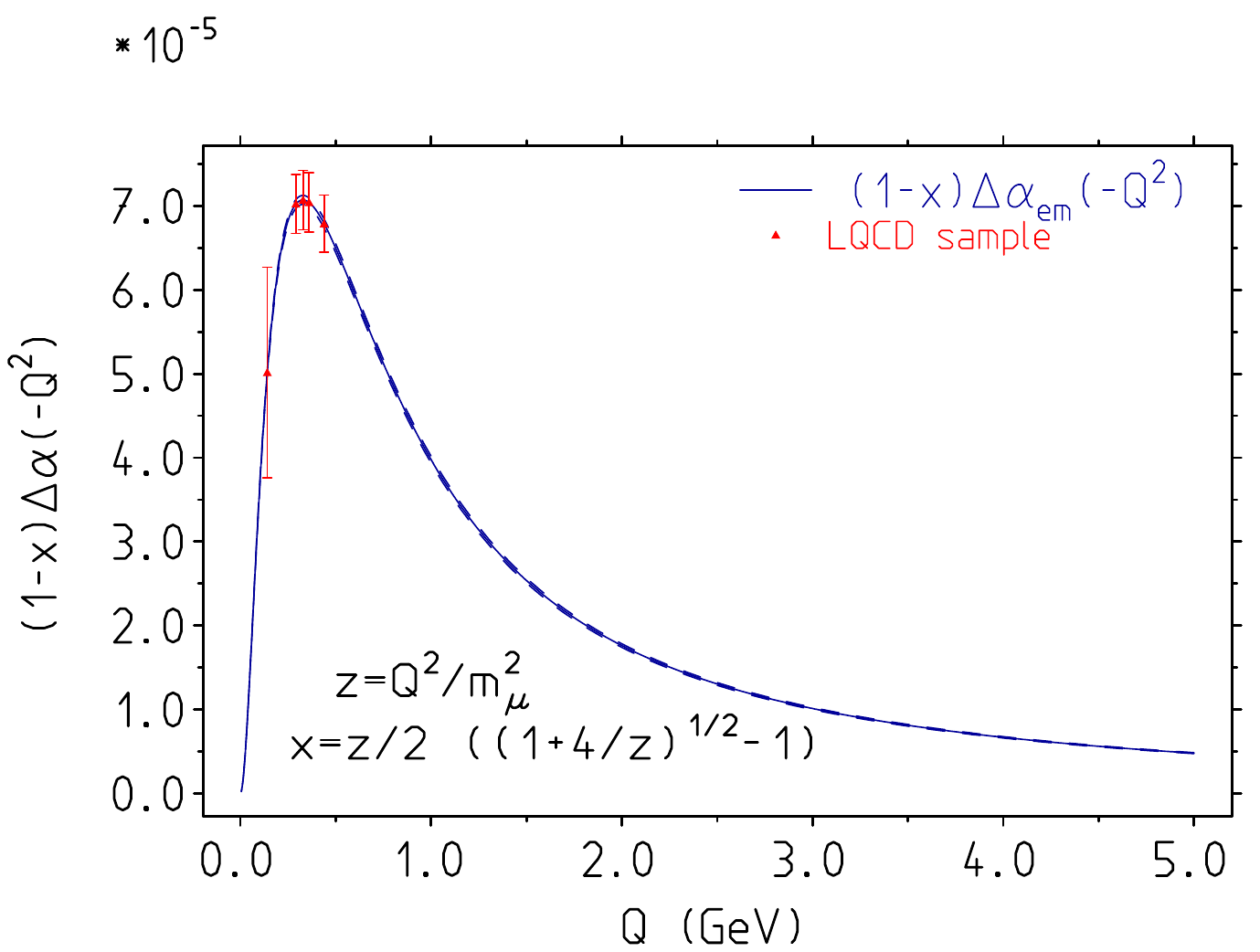}
\caption{The integrand of the vacuum polarization representation (\ref{RAI}) as a function of $x$ and as a
function of the energy scale $Q$. As we see the integrand is strongly
peaked as a function of $Q$ at about $330~\mv$. $\Pi (Q^2)$ data come
from~\cite{hvpfunction}.  The dashed lines mark the error band from
the experimental data.  ``LQCD sample'' as in
Fig.~\ref{fig:ADIkernels}.  In the left panel we again display
the contributions to $\amuh$ from
regions between $Q_i=0.00,\,0.15,\,0.30,\,0.45$ and $1.0~\gv$ in
percent. The tail above 1~GeV contributes slightly less than 1\%. Note
the different distribution of the contributions from the different
ranges for the Adler function integral representation 
(see left panel of Fig.~\ref{fig:ADIkernels}).}
\label{fig:RAIkernels} 
\end{figure}

In this section we study the Mellin-Barnes moment (MBM) approach in terms of Euclidean
quantities as proposed in~\cite{deRafael:2014gxa}.
The phenomenologically estimated moments provide useful tests for
lattice results since the moments
\be
\cM(-n)=\frac{(-1)^{(n+1)}}{(n+1)!}\,\left(m_\mu^2\right)^{n+1}\:
\left.\left(\frac{\partial^{n+1}}{\left(\partial Q^2\right)^{n+1}}\,\Pi(Q^2)\right)\right|_{Q^2=0}\epo
\label{basicmoments}
\ee
are directly accessible by lattice QCD.  
A comparison with the Taylor expansion
\be
\Pi(Q^2)=\sum_{n=0}^{\infty}\,\left(Q^2\right)^{(n+1)}\,\frac{1}{(n+1)!} 
\left.\left(\frac{\partial^{n+1}}{\left(\partial Q^2\right)^{n+1}}\,\Pi(Q^2)\right)\right|_{Q^2=0}=
\sum_{n=0}^{\infty}\,\left(Q^2\right)^{(n+1)}\,\Pi_{n+1}\,,
\label{Taylor}
\ee
reveals that the moments $\cM(-n)$, up to normalization, agree with the
normal Taylor coefficients of the low energy expansion of $\Pi(Q^2)$:
\be
\Pi_{n+1}=(-1)^{(n+1)}\,\left(m_\mu^2\right)^{-(n+1)}\,\cM(-n)\,.
\label{Taylorcoefs}
\ee
Up to a factor $4\pi\alpha Q_f^2$ summed over the $n_f=4$ flavors $f$
included, these are the moments used in the recent
analysis~\cite{Chakraborty:2016mwy}, for example\footnote{The
uncorrected Taylor coefficients
for sets 8 and 10 (the closest to the physical point) of Table~II in \cite{Chakraborty:2016mwy} 
translate into $\cM(-n)=4\pi\alpha\,\sum
Q_f^2\,(-1)^n\,(m^2_\mu)^{n+1}\,\Pi_{n+1}$ as follows:\\[-4mm]
\begin{center}
{\scriptsize
\begin{tabular}{cclcclcclccl}
\hline \hline
 $\Pi_1$&=&$ 0.0811(12) $&
 $\Pi_2$&=&$ 0.1238(36) $&
 $\Pi_3$&=&$ 0.205(9)   $&
 $\Pi_4$&=&$ 0.344(20)  $\\
  $\cM(0) $&=&$  9.23(14)    $&
  $\cM(-1)$&=&$  0.1572(46)	 $&
  $\cM(-2)$&=&$  0.00291(13) $&
  $\cM(-3)$&=&$  0.000054(3) $\\
\hline
$\Pi_1 $&=&$ 0.0756(13) $&
$\Pi_2 $&=&$ 0.1111(41) $&
$\Pi_3 $&=&$ 0.179(11)  $&
$\Pi_4 $&=&$ 0.293(25)  $\\
 $ \cM(0) $&=&$  8.60(15)    $&
 $ \cM(-1)$&=&$  0.1411(52)	 $&
 $ \cM(-2)$&=&$  0.00254(16) $&
 $ \cM(-3)$&=&$  0.000046(4) $\\
\hline
\end{tabular}
}	
\end{center}	  
to be compared with the HLS model column of Table~\ref{tab:mom}. 
Note
that substantial corrections to be applied to the raw data are not included here.}. 
A partial integration allows us
to rewrite (\ref{ADI}) directly as an integral over the vacuum
polarization amplitude~\cite{LPdR72,Jegerlehner:2008zza}
\be
\amuh =\frac{\alpha}{\pi}\int\limits_0^1 \D x\:(1-x)\: \dalh
\left(-Q^2(x)\right)=-\frac{\alpha}{\pi}\,\int\limits_0^1 \D x\:(1-x)\:\Pi
\left(Q^2(x)\right)
\label{RAI}
\ee
and inserting the Taylor expansion for $\Pi(Q^2(x))$, with $Q^2(x)\equiv \frac{x^2}{1-x}m_\mu^2$  we get
\be
\amuh= \frac{\alpha}{\pi}\,\sum_{n=0}^{\infty}\,(-1)^n\,\cM(-n)\,
\int\limits_0^{x_1} \D x\:x^2\,\left(\frac{x^2}{1-x}\right)^n
-\frac{\alpha}{\pi}\,\int\limits_{x_1}^1 \D x\:(1-x)\:\Pi
\left(Q^2(x)\right)\,,
\label{RAITaylor}
\ee
which requires an appropriate energy cutoff $x_1 < 1$ at which the low
momentum expansion ceases to make sense. Obviously, the expansion
collapses for an upper limit $x_1=1$. Not surprisingly, the problem
is the high energy tail; an Euclidean cutoff $Q^2_1$ indeed provides
an effective $x_1=\frac{q^2_1}{2}\,\sqrt{1+4/q^2_1}-1 \approx 1-1/q^2_1
+ \cdots$ where $q_1=Q_1/m_\mu$. Here we are confronted with the
question about the dependence of the result on the cutoff. This is
different for the timelike representation (\ref{lohadalt}), due to
the $1/s^2$ behavior of the kernel, while $R(s)$ approaches a
constant. The cutoff dependence is suppressed by $1/E^2_1$ for high
enough cutoffs $E_1$ in this case. In order to learn where the
dominant contributions come from,  we plot the integrand of (\ref{RAI})
in Fig.~\ref{fig:RAIkernels}. Also in the Euclidean region the
integrand is highly peaked, now around half of the $\rho$ meson mass
scale.

\begin{figure}
\centering
\includegraphics[width=0.65\textwidth]{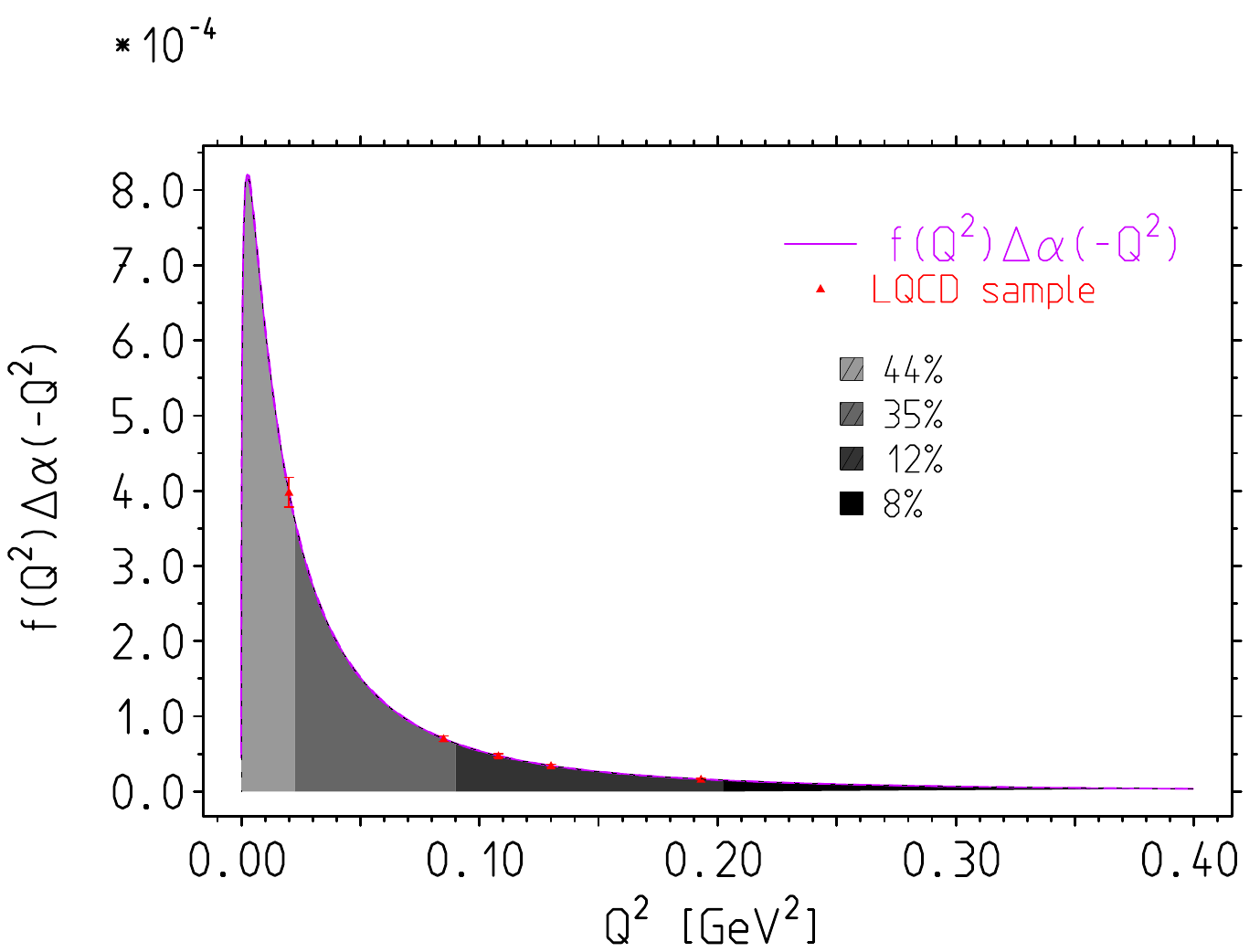}
\caption{The integrand of (\ref{LQCDI}), which represents (\ref{RAI}) as an integral over
$Q^2$. Ranges between $Q_i=0.00,\,0.15,\,0.30,\,0.45$ and $1.0~\gv$
and their percent contibution to $\amuh$ and the ``LQCD sample'' as in
Fig.~\ref{fig:RAIkernels}.}
\label{fig:LQCDkernels} 
\end{figure}

Lattice QCD groups usually use
a different representation for the $\amuh$ dispersion integral:
\ba
\amuh [Q^2_{\rm max}]= \frac{\alpha}{\pi}\,\int_0^{Q^2_{\rm max}} dQ^2
\,f(Q^2)\,(-4\pi\alpha\hat{\Pi}(Q^2))\,,
\label{LQCDI}
\ea
with 
$$f(Q^2)=m_\mu^2 Q^2Z^3(Q^2)\,(1-Q^2 Z(Q^2))/(1+m_\mu^2 Q^2 Z^2(Q^2))$$ and
$$Z(Q^2)=\left(\sqrt{Q^4+4m_\mu^2Q^2}-Q^2\right)/(2m_\mu^2Q^2)\,.$$ In
our notation $-4\pi\alpha\hat{\Pi}(Q^2)=\Delta \alpha_{\rm had}(-Q^2)\,.$ See
e.g. Figure 1 of ~\cite{Aubin:2015rzx} which in our representations
(\ref{RAI}) and (\ref{ADI}) translates into our Figs.~\ref{fig:RAIkernels}
and~\ref{fig:ADIkernels}, respectively. The contributions to $\amuh$ from the
ranges displayed in the left panel of Fig.~\ref{fig:RAIkernels} are
the same. This is illustrated in Fig.~\ref{fig:LQCDkernels}. 
We note that in the representation
(\ref{ADI}), in terms of the Adler function, the
contribution obtained for a given $Q_{\rm min}$ is substantially
smaller than in the representations (\ref{RAI}) or (\ref{LQCDI}), which integrate the
HVP function directly.
  
\begin{figure}[h]
\centering
\includegraphics[height=7cm]{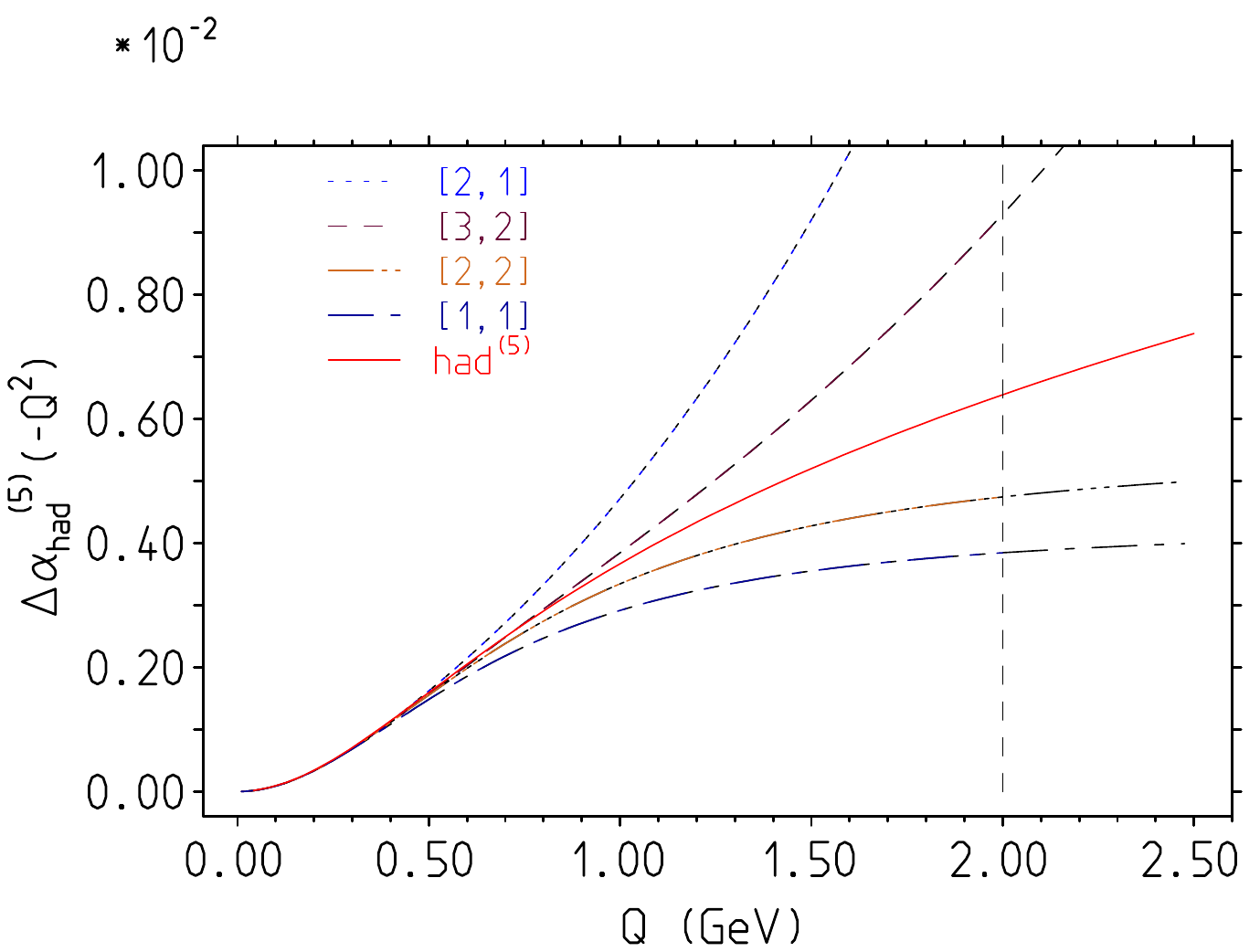}
\caption{$\dalh$ as a function of the spacelike momentum transfer $Q$
together with the best pairs of Pad\'e approximants ([1,1] and [2,1])
and ([2,2] and [3,2]), which can be formed given 4 and 5 moments
$\cM(-n)$ ($n=0,1,2,3,4$), respectively. Keep in mind that the
reference scale of our moment expansion is the muon mass $m_\mu$ and
higher momenta get suppressed in the $\amuh$ integral. The ``exact'' $\dalh$
labeled by {\tt had$^{(5)}$} comes from~\cite{hvpfunction}.}
\label{fig:HVPPades} 
\end{figure}

In the lattice QCD approach, the current correlator defining $\Pi(Q^2)$
is evaluated in configuration space, and one would have to perform a
Fourier transformation, which, for obvious reasons, is not so straightforward 
with the discrete lattice data. The moments $\Pi_n$ of
Eq.~(\ref{Taylorcoefs}), on the other hand are directly
accessible by calculating 
\be
G_{2j}\equiv \sum_t \sum_{\vec{x}}t^{2j}Z_V^2 \langle
j^i(\vec{x},t)j^i(0) \rangle
=\left. (-1)^j\,\frac{\partial ^{2j}}{\partial k^{2j}}\,k^2 \hat{\Pi}(k^2)\right|_{k^2=0}
=(-1)^j\,(2j!)\Pi_{j-1}\,,
\label{latticeTayCoe}
\ee
where $Z_V$ is the lattice vector current renormalization factor and
the sums extend over the time $t$ and space $\vec{x}$ lattice points.
So, what is available primarily is the low momentum expansion only. In
order to get a useful $\Pi(Q^2)$ for the higher momenta, one usually
calculates the Pad\'e approximants~\cite{Chakraborty:2016mwy}, which
then allow for an acceptable estimate of the full contribution.  In
Fig.~\ref{fig:HVPPades} we show results in comparison with $\dalh$.
Best bounds are the Pad\'es of type [n,n] as lower bound accompanied
by [n+1,n] as upper constraint, which requires $2n+1$
coefficients\footnote{The number of coefficients of a Pad\'e
approximant [m,n]=$\sum_{k=0}^{m}\,a_k x^k/(1+\sum_{k=1}^{n}\,b_kx^k)$
is $n+m+1$ unless $a_0=0$ as in case of the HVP $\Pi(Q^2)$ or the
Adler function $D(Q^2)$, where it is $n+m$. Pad\'es for $D(Q^2)/Q^2$
and for the truncated HVP $\Pi(Q^2)_n^{\rm trunc}/(Q^2)^{n+1}$
considered in Sect.~\ref{sec:LQCDhowto} require $n+m+1$ coefficients,
however.}. For the five moments we have worked out, adopting the
moments ``HLS+remainder'', the best Pad\'e approximants are [1,1] and
[2,1] requiring 3 moments, [2,1] and [2,2] requiring 4 moments and
[2,2] and [3,2] for given 5 moments. The integrals of the Pad\'es are
listed in Table~\ref{tab:TaylorPades} together with the result from the
direct integration (DR) and from the MBM expansion of order $n=4$
(five moments $\cM(-n)$ and four moments $\tM(-n)$). The upper bound
Pad\'es [2,1] and [3,2] are combined with the lower bound ones [1,1]
and [2,2], respectively, taking half of the sums and adding half of
the difference as a model error in quadrature. The different results
are in good agreement with each other.
\begin{table}[h]
\centering
\caption{$\amuh \power{10}$ from the Pad\'e approximants of the $\Pi(Q^2)$ Taylor
expansion in $Q^2/m_\mu^2$, given 3, 4 and 5 coefficients of the ``HLS + remainder''
moments.}
\label{tab:TaylorPades}
\begin{tabular}{cc||cc||cc}
\noalign{\smallskip}\hline\noalign{\smallskip}
\multicolumn{2}{c}{3 Taylor coefficients} & \multicolumn{2}{c}{4 Taylor coefficients}& \multicolumn{2}{c}{5 Taylor coefficients} \\
\noalign{\smallskip}\hline\noalign{\smallskip}
\mbox{[1,1]} & $672.21 \pm \phantom{1}4.16$ & \mbox{[2,1]} & $688.96 \pm 4.24$& \mbox{[2,2]} & $679.13 \pm 4.20$\\
\mbox{[2,1]} & $688.96 \pm \phantom{1}4.24$ & \mbox{[2,2]} & $679.13 \pm 4.20$& \mbox{[3,2]} & $682.23 \pm 4.22$\\
\mbox{[1,1]}+\mbox{[2,1]} & $680.58 \pm \phantom{1}9.37$&\mbox{[2,1]}+\mbox{[2,2]} & $684.04 \pm 6.48$&\mbox{[2,2]}+\mbox{[3,2]} & $680.68 \pm 4.49$\\
\noalign{\smallskip}\hline\noalign{\smallskip}
\multicolumn{6}{c}{DR\phantom{xxx}$681.77 \pm \phantom{1}3.14$\phantom{xxx}  MBM \phantom{xxx} $681.48 \pm 4.18$}  \\
\noalign{\smallskip}\hline
\end{tabular}
\end{table}
We have used that the factor $(1-x)$ in (\ref{RAITaylor}) acts as a
$1/Q^2$ factor at high energy, such that the [n+1,n] Pad\'es are not
in conflict with integrability. Nevertheless, the Pad\'es cannot be
arranged to be in accord with QCD asymptotics and a suitable
modification taking into account this fact is appropriate. In the
Euclidean region the ``experimental'' Adler function can be used to
check the validity of pQCD~\cite{EJKV98}. 
One finds that pQCD works
pretty accurately above about 2~GeV to 2.5~GeV. We have adopted a
cutoff of $Q_1=2~\gv$ in order to obtain the results in Fig.~\ref{fig:HVPPades}
and Table~\ref{tab:TaylorPades}, where we use $\dalh(-Q^2)$ for momenta
$Q>Q_1$. In fact the results presented do not substantially depend on
the cut  and remain within uncertainties.

\begin{figure}[h]
\centering
\includegraphics[height=7cm]{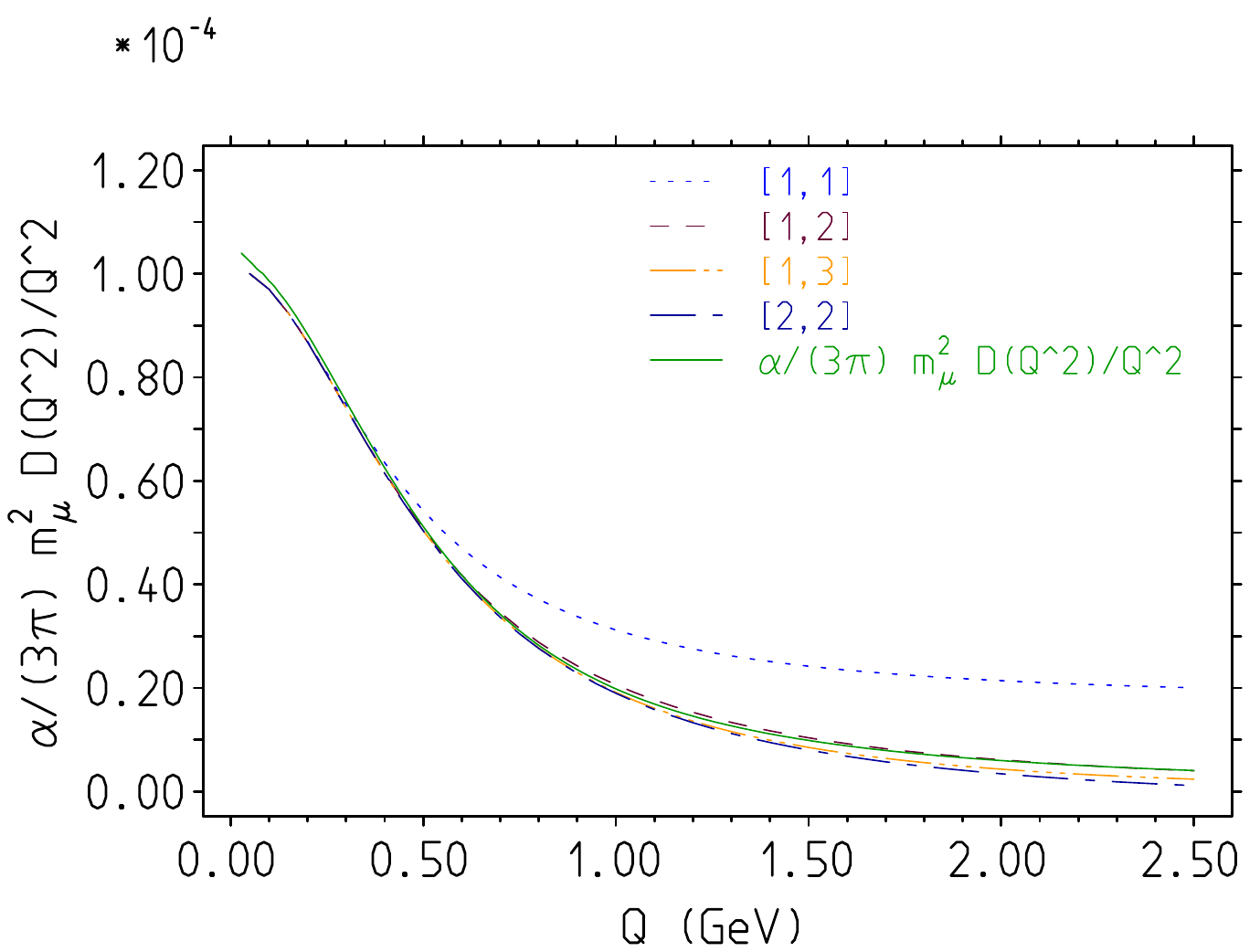}
\caption{Comparison of the best Pad\'e approximants for the Taylor
expansion of $\frac{\alpha}{3\pi}\,m_\mu^2\,D(Q^2)/Q^2$, which enters the representation
(\ref{ADI}). The best pair of Pad\'es for our
set of 5 coefficients are [1,2] as upper and [1,3] (or [2,2]) as lower
bound. Most of the Pad\'es fail to represent the data towards higher energies.
The ``exact'' Adler function ratio $D(Q^2)/Q^2$ comes from~\cite{Adlerfunction}.}
\label{fig:ADPPades}
\end{figure}

Similarly, we may look at the ``Taylor + Pad\'e'' method for the Adler
function representation (\ref{ADI}).  From (\ref{Taylor}) and
(\ref{Taylorcoefs}), we learn that, given the HVP function Taylor coefficients
\be
\Pi(Q^2)=\sum_{n=0}^{\infty}\,\left(Q^2\right)^{(n+1)}\,(-1)^{(n+1)}\,\left(m_\mu^2\right)^{-(n+1)}\,\cM(-n)\,,
\label{TaylorII}
\ee
the corresponding Adler function ones follow by the replacement
$\cM(-n) \to (n+1)\,\cM(-n)$ as 
\be
D(Q^2)\propto - Q^2 \frac{\D }{\D Q^2}\,\Pi(Q^2)= \sum_{n=0}^{\infty}\,\left(Q^2\right)^{(n+1)}\,(-1)^{n}\,\left(m_\mu^2\right)^{-(n+1)}\,(n+1)\,\cM(-n)\,.
\ee
Equations  (\ref{ADI}) and (\ref{DD}) suggest to consider
\be
\hat{\cal D}(Q^2)\equiv \frac{\alpha}{3\pi}\,m_\mu^2\,D(Q^2)/Q^2
\ee
and expand in $Q^2/m_\mu^2$. The corresponding best Pad\'es are shown
in Fig.~\ref{fig:ADPPades}, where ``best'' cases are chosen such that
they match best the ``experimental'' curve, extracted by means of
(\ref{DI}) from $\epm$ data, towards higher energies.
Note that the Pad\'es for $\hat{\cal D}(Q^2)$ differ from those of the
HVP function as, by definition, its Taylor expansion starts with a constant term.
Given 5 coefficients, the best estimate is given in
Table~\ref{tab:TaylorPadesAdler} by the pair [1,2] and [1,3].
\begin{table}[h]
\centering
\caption{$\amuh \power{10}$ from the Pad\'e approximants of the Taylor
expansion of $D(Q^2)/Q^2$, given 5 coefficients of the ``HLS + remainder''
moments.}
\label{tab:TaylorPadesAdler}
\begin{tabular}{cc|cc|cc}
\noalign{\smallskip}\hline\noalign{\smallskip}
\mbox{[1,2]} & $681.35 \pm 4.21$ &
\mbox{[1,3]} & $680.74 \pm 4.25$ &
\mbox{[1,2]}+\mbox{[1,3]} & $681.05 \pm 4.24$\\
\noalign{\smallskip}\hline
\end{tabular}
\end{table}

In contrast to the Taylor expansion approach, the Mellin-Barnes
representation provides a convergent expansion in terms of the moments
$\cM(-n)$ and $\tilde{\cM}(-n)$.  The log weighted moments
$\tilde{\cM}(-n)$ are related to properties of $\Pi(Q^2)$ in a more
complicated manner~\cite{deRafael:2014gxa} and require, in addition, to
determine the moments
\be
\Sigma(-n;Q_0^2)\equiv \int\limits_{Q_0^2}^{\infty}\,\D Q^2\,\left(\frac{m_\mu^2}{Q^2}\right)^{n+1}\:
\left(-\frac{\Pi(Q^2)}{Q^2}\right)
\label{sigma}
\ee
for $n=1,2,3,\cdots$. Note that there is here an ambiguity in the
choice of $Q_0^2$ as these are not integrals along a cut, as the
integrals over $R(s)$ are. Nevertheless, we need $Q_0^2>0$ to provide an
infrared cutoff, in order for the low momentum expansion moments to exist.
As suggested in~\cite{deRafael:2014gxa} we may adopt the choice
$Q_0^2=4\,m_\pi^2$. However, $Q_0^2=m_{\pi^0}^2$ may be a better
choice as will seen below, where we compare the two choices
\be
Q_0^2=m_{\pi^0}^2~ \mathrm{ \ and \ }~ Q_0^2=4\,m_\pi^2\epo
\label{Euclideanthreshold}
\ee
The relation between $\tilde{\cM}(-n)$ and $\Sigma(-n; s_0)$, which also
involves the moments $\cM(-n)$, can be found by means of applying the
subtracted dispersion relation:
\be
-\frac{\Pi(Q^2)}{Q^2}=\frac{\alpha}{3\pi}\,\int\limits_{4m_\pi^2}^{\infty}\,\frac{\D s}{s}\,\frac{R(s)}{s+Q^2}\,\epo
\label{HVPDR}
\ee
We note that the Euclidean moments $\Sigma(-n; s_0)$ in terms of $R(s)$
appear represented as double integrals, where the first integration
transforms the timelike $R(s)$ information into the spacelike vacuum
polarization function $\Pi(Q^2)$, a smoothed object, devoid of
thresholds and of resonance peaks. However, interchanging integrations,
the first integration represents a kernel $J(s_0,s;n)$ which, for each
$n$, can be performed analytically. Therefore, also in this case, one
ends up with a one-dimensional integral representation.  For
$\Sigma(-n; s_0)$
in terms of moments one obtains:
\ba
\label{Sigmadire}
\Sigma(-n;s_0)&=&\frac{\alpha}{3\pi}\,\int\limits_{4 m_\pi^2}^{\infty}\,\frac{\D s}{s}\,R(s)\,J(s_0,s;n)\\
J(s_0,s;n)&=&\int\limits_{Q^2_0=s_0}^{\infty}\,\D Q^2 \left(\frac{m_\mu^2}{Q^2}\right)^{n+1}\:\frac{1}{s+Q^2}\nn
\epo
\ea
The second integral can be performed analytically and up to 3$^{\rm
rd}$ order yields:
\ba
J(s_0,s;1) &=&-\ln(1+\frac{s}{s_0})\,\left(\frac{m_\mu^2}{s}\right)^2+\frac{m_\mu^2}{s_0}\frac{m_\mu^2}{s}\crn
J(s_0,s;2) &=& \ln(1+\frac{s}{s_0})\,\left(\frac{m_\mu^2}{s}\right)^3-\frac{m_\mu^2}{s_0}\left(\frac{m_\mu^2}{s}\right)^2
+\frac12\,\left(\frac{m_\mu^2}{s_0}\right)^2\,\frac{m_\mu^2}{s}
\crn
J(s_0,s;3) &=&-\ln(1+\frac{s}{s_0})\,\left(\frac{m_\mu^2}{s}\right)^4+\frac{m_\mu^2}{s_0}\left(\frac{m_\mu^2}{s}\right)^3
-\frac12\,\left(\frac{m_\mu^2}{s_0}\right)^2\,\left(\frac{m_\mu^2}{s}\right)^2
\crn &&+\frac13\,\left(\frac{m_\mu^2}{s_0}\right)^3\,\frac{m_\mu^2}{s}\epo 
\label{Jkern}
\ea
Then, using
$ \displaystyle
\ln(1+\frac{s}{s_0})=-\ln(\frac{m_\mu^2}{s})-\ln(\frac{s_0}{m_\mu^2})+\ln
\left(1+\frac{s_0}{s}\right)$, we obtain:
\ba
\Sigma(-1;s_0) &=&\tilde{\cM}(-1)+\ln \frac{s_0}{m_\mu^2}\,\cM(-1)+ \frac{m_\mu^2}{s_0}\,\cM(0)
-\cR(-1;s_0)
\crn
\Sigma(-2;s_0) &=&
-\tilde{\cM}(-2)-\ln \frac{s_0}{m_\mu^2}\,\cM(-2)- \frac{m_\mu^2}{s_0}\,\cM(-1)+\frac12\,\left(\frac{m_\mu^2}{s_0}\right)^2\,\cM(0)
+\cR(-2;s_0)
\crn
\Sigma(-3;s_0) &=&
\tilde{\cM}(-3)+\ln \frac{s_0}{m_\mu^2}\,\cM(-3)+ \frac{m_\mu^2}{s_0}\,\cM(-2)
-\frac12\,\left(\frac{m_\mu^2}{s_0}\right)^2\,\cM(-1)\crn && +\frac13\,\left(\frac{m_\mu^2}{s_0}\right)^3\,\cM(0)
-\cR(-3;s_0)
\label{sigmafrommoms}
\ea
with the ``remainder'':
\be
\cR(-n;s_0)= \frac{\alpha}{3\pi}\,\int\limits_{4 m_\pi^2}^{\infty}\,\frac{\D
s}{s}\,R(s)\,\ln \left(1+\frac{s_0}{s}\right)\,\left(\frac{m_\mu^2}{s}\right)^{n+1}\epo
\label{Rdire}
\ee
The latter can be evaluated in terms of $e^+e^-$ data  and BHLS
predictions, but  they are not directly accessible by lattice
data.
However, in this representation the log can be expanded as
$\ln(1+x)=\sum_{n=1}^{\infty} (-1)^{n+1}\,\frac{x^n}{n}$ which converges for $-1 <x\leq
1$ and we obtain a series of normal moments $\cM(-n)$ accessible by
LQCD. We thus have
\be
\cR(-n;s_0)\approx \frac{s_0}{m_\mu^2}\,\cM(-n-1) -\frac12\,\left(\frac{s_0}{m_\mu^2}\right)^2\,\cM(-n-2) +\cdots
\label{Rapprox}
\ee
Here we see that the choice of $s_0$, or $Q_0^2$ in (\ref{sigma}),
respectively, is not uncritical, if we want the series to converge
well. For the isovector part $0<s_0\leq 4 m_{\pi}^2$ is adequate. For
the full electromagnetic case, with $m_{\pi^0}$ being the true
threshold $0<s_0\leq m_{\pi^0}^2$ is appropriate. The results are
presented in Table~\ref{tab:rem}. While the approximate results
$\cR^{\approx}(-n; s_0)$ agree fairly well  for $s_0=m^2_{\pi^0}$
with the direct evaluations $\cR^*(-n; s_0)$, for $s_0=4 m^2_{\pi}$ the
agreement is not convincing. One has to keep in mind that
$\amuh$ evaluations are required with high precision.
\begin{table}
\centering
\caption{The remainders $\cR(-n;s_0)$ in
units $10^{-5}$  for $s_0=m_{\pi^0}^2$ in the upper part and for 
$s_0=4 m_{\pi}^2$ in the lower part. The $\cR^*(-n;s_0)$ are evaluated
in via (\ref{Rdire}) the same way as the moments of
Table~\ref{tab:mom}. The remainders $\cR^{\approx}(-n;s_0)$ are calculated
via the expansion (\ref{Rapprox}) in terms of the moments $\cM(-n)$
from Table~\ref{tab:mom}}
\label{tab:rem}
{\scriptsize
\begin{tabular}{lr@{.}lr@{.}l|r@{.}lr@{.}l|r@{.}lr@{.}l|r@{.}lr@{.}l}
\noalign{\smallskip}\hline\noalign{\smallskip}
moment &
\multicolumn{4}{c}{data direct} &
\multicolumn{4}{c}{data HLS channels} &
\multicolumn{4}{c}{HLS model} &
\multicolumn{4}{c}{HLS + remainder}  \\
\noalign{\smallskip}\hline\noalign{\smallskip}
$\cR^*(-1;m_{\pi^0}^2) $
& 0&013602&$\!\!\!\!\!\pm$0&000177
& 0&013549&$\!\!\!\!\!\pm$0&000177
& 0&014010&$\!\!\!\!\!\pm$0&000018
& 0&014063&$\!\!\!\!\!\pm$0&000020   \\
$\cR^*(-2;m_{\pi^0}^2) $
& 0&0007473&$\!\!\!\!\!\pm$0&0000142
& 0&0007469&$\!\!\!\!\!\pm$0&0000142
& 0&0007920&$\!\!\!\!\!\pm$0&0000011
& 0&0007924&$\!\!\!\!\!\pm$0&0000011\\
$\cR^*(-3;m_{\pi^0}^2) $
& 0&00005591&$\!\!\!\!\!\pm$0&00000126
& 0&00005591&$\!\!\!\!\!\pm$0&00000126
& 0&00006015&$\!\!\!\!\!\pm$0&00000008
& 0&00006016&$\!\!\!\!\!\pm$0&00000008 \\
\noalign{\smallskip}\hline
$\cR^{\approx}(-1;m_{\pi^0}^2) $
& 0&013609&$\!\!\!\!\!\pm$0&000177
& 0&013556&$\!\!\!\!\!\pm$0&000177
& 0&014018&$\!\!\!\!\!\pm$0&000017
& 0&014071&$\!\!\!\!\!\pm$0&000017   \\
$\cR^{\approx}(-2;m_{\pi^0}^2) $
& 0&0007429&$\!\!\!\!\!\pm$0&0000141
& 0&0007426&$\!\!\!\!\!\pm$0&0000141
& 0&0007873&$\!\!\!\!\!\pm$0&0000010
& 0&0007877&$\!\!\!\!\!\pm$0&0000010  \\
$\cR^{\approx}(-3;m_{\pi^0}^2) $
& 0&00005999&$\!\!\!\!\!\pm$0&00000135
& 0&00005999&$\!\!\!\!\!\pm$0&00000135
& 0&00006456&$\!\!\!\!\!\pm$0&00000008
& 0&00006456&$\!\!\!\!\!\pm$0&00000008  \\
\noalign{\smallskip}\hline\noalign{\smallskip}
\noalign{\smallskip}\hline
$\cR^*(-1;4 m_{\pi}^2) $
& 0&0517876 &$\!\!\!\!\!\pm$0&0006421
& 0&0515651 &$\!\!\!\!\!\pm$0&0006421
& 0&0531340 &$\!\!\!\!\!\pm$0&0000674   
& 0&053357  &$\!\!\!\!\!\pm$0&0000674 \\
$ \cR^*(-2;4 m_{\pi}^2) $
& 0&0027449 &$\!\!\!\!\!\pm$0&0000508
& 0&0027433 &$\!\!\!\!\!\pm$0&0000508
& 0&0028909 &$\!\!\!\!\!\pm$0&0000037   
& 0&0028925 &$\!\!\!\!\!\pm$0&0000037  \\
$  \cR^*(-3;4 m_{\pi}^2) $
& 0&00019963&$\!\!\!\!\!\pm$0&00000448
& 0&00019962&$\!\!\!\!\!\pm$0&00000448
& 0&00021152&$\!\!\!\!\!\pm$0&00000027  
& 0&00021152&$\!\!\!\!\!\pm$0&00000027 \\
\noalign{\smallskip}\hline
$ \cR^{\approx}(-1;4 m_{\pi}^2) $
& 0&053085 &$\!\!\!\!\!\pm$0&000670
& 0&052860 &$\!\!\!\!\!\pm$0&000670
& 0&054582 &$\!\!\!\!\!\pm$0&000067
& 0&054808 &$\!\!\!\!\!\pm$0&000073   \\
$ \cR^{\approx}(-2;4 m_{\pi}^2) $
& 0&0024911 &$\!\!\!\!\!\pm$0&0000447
& 0&0024897 &$\!\!\!\!\!\pm$0&0000447
& 0&0026288 &$\!\!\!\!\!\pm$0&0000032  
& 0&0026302 &$\!\!\!\!\!\pm$0&0000435 \\
$ \cR^{\approx}(-3;4 m_{\pi}^2) $
& 0&0002566&$\!\!\!\!\!\pm$0&0000058
& 0&0002566&$\!\!\!\!\!\pm$0&0000058     
& 0&00027612&$\!\!\!\!\!\pm$0&00000035
& 0&00027612&$\!\!\!\!\!\pm$0&00000035  \\
\noalign{\smallskip}\hline
\end{tabular}
}
\end{table}

Knowing the remainders $\cR(-n;s_0)$, either by direct integration of
(\ref{Rdire}) or using their expansion in terms of moments $\cM(-n)$
we are able to calculate the Euclidean moments $\Sigma(-n;s_0)$ via $R(s)$
data or BHLS model predictions with the help of the relations
(\ref{sigmafrommoms}) in terms of the moments and remainders given in
Table~\ref{tab:mom} and Table~\ref{tab:rem},
respectively. Table~\ref{tab:sig} lists the results of this
evaluation. As mentioned earlier, the moments $\Sigma(-n;s_0)$ are
very much dependent on the choice of $s_0$, which actually should not
exceed the threshold $m_{\pi^0}^2$, unless we restrict the analysis to
the isovector part with threshold at $4 m_{\pi}^2$. Note that
actually, the $\cR(-n,s_0)$'s vanish for $s_0\to 0$. So, if we choose
$s_0$ small enough, the $\cR(-n,s_0)$'s can be tuned to be negligible.
\begin{table}
\centering
\caption{The moments $\Sigma(-n; s_0)$ in
units $10^{-5}$ for $s_0=m_{\pi^0}^2$ in the upper part and for $s_0=4
m_{\pi}^2$ in the lower part. By $\Sigma^*(-n; s_0) $ we denote the result
from (\ref{Sigmadire}) and the kernels (\ref{Jkern}). The version
$\Sigma(-n; s_0)$ 
denotes the result of (\ref{sigmafrommoms}) using the ``exact'' remainder
$\cR^*(-n;s_0)$ while $\Sigma^{\approx}(-n; s_0) $ is the result obtained by
the truncated expansion (\ref{Rapprox}) including moments up to $n=4$.}
\label{tab:sig}
{\scriptsize
\begin{tabular}{lr@{.}lr@{.}l|r@{.}lr@{.}l|r@{.}lr@{.}l|r@{.}lr@{.}l}
\noalign{\smallskip}\hline\noalign{\smallskip}
moment &
\multicolumn{4}{c}{data direct} &
\multicolumn{4}{c}{data HLS channels} &
\multicolumn{4}{c}{HLS model} &
\multicolumn{4}{c}{HLS + remainder}  \\
\noalign{\smallskip}\hline\noalign{\smallskip}
$\Sigma^*(-1;m_{\pi^0}^2) $
&  5&46769&$\!\!\!\!\!\pm$0&04183
&  4&58924&$\!\!\!\!\!\pm$0&02570
&  4&57130&$\!\!\!\!\!\pm$0&0070
&  5&44975&$\!\!\!\!\!\pm$0&0337   \\
$\Sigma^*(-2;m_{\pi^0}^2) $
&  1&77639&$\!\!\!\!\!\pm$0&01335
&  1&50224&$\!\!\!\!\!\pm$0&00846
&  1&4969&$\!\!\!\!\!\pm$0&0023
&  1&77105&$\!\!\!\!\!\pm$0&0106 \\
$\Sigma^*(-3;m_{\pi^0}^2) $
&  0&735060&$\!\!\!\!\!\pm$0&005500
&  0&622759&$\!\!\!\!\!\pm$0&003517
&  0&62063&$\!\!\!\!\!\pm$0&00095
&  0&732931&$\!\!\!\!\!\pm$0&0043 \\  
\noalign{\smallskip}\hline                	
$\Sigma(-1;m_{\pi^0}^2) $  			
&  5&48300&$\!\!\!\!\!\pm$0&04122
&  4&58905&$\!\!\!\!\!\pm$0&02604
&  4&57106&$\!\!\!\!\!\pm$0&00697
&  5&46501&$\!\!\!\!\!\pm$0&03350  \\
$\Sigma(-2;m_{\pi^0}^2) $
&  1&78098&$\!\!\!\!\!\pm$0&01310
&  1&50210&$\!\!\!\!\!\pm$0&00842
&  1&49675&$\!\!\!\!\!\pm$0&00228
&  1&77563&$\!\!\!\!\!\pm$0&01054 \\ 
$ \Sigma(-3;m_{\pi^0}^2) $
&  0&73690&$\!\!\!\!\!\pm$0&00542
&  0&62267&$\!\!\!\!\!\pm$0&00350
&  0&62053&$\!\!\!\!\!\pm$0&00094
&  0&73476&$\!\!\!\!\!\pm$0&00432 \\ 
\noalign{\smallskip}\hline    			 
$\Sigma^{\approx}(-1;m_{\pi^0}^2) $
&  5&48300&$\!\!\!\!\!\pm$0&04122
&  4&58905&$\!\!\!\!\!\pm$0&02605
&  4&57106&$\!\!\!\!\!\pm$0&00697
&  5&46501&$\!\!\!\!\!\pm$0&03352   \\ 
$\Sigma^{\approx}(-2;m_{\pi^0}^2) $
&  1&78098&$\!\!\!\!\!\pm$0&01310
&  1&50210&$\!\!\!\!\!\pm$0&00842
&  1&49675&$\!\!\!\!\!\pm$0&00228
&  1&77562&$\!\!\!\!\!\pm$0&01054 \\ 
$\Sigma^{\approx}(-3;m_{\pi^0}^2) $
&  0&73690&$\!\!\!\!\!\pm$0&00542
&  0&62267&$\!\!\!\!\!\pm$0&00350
&  0&62052&$\!\!\!\!\!\pm$0&00094
&  0&73475&$\!\!\!\!\!\pm$0&00432 \\	 
\noalign{\smallskip}\hline\noalign{\smallskip}
\noalign{\smallskip}\hline
$ \Sigma^*(-1;4 m_{\pi}^2) $
&  1&02683&$\!\!\!\!\!\pm$0&00852
&  0&83419&$\!\!\!\!\!\pm$0&00457
&  0&82971&$\!\!\!\!\!\pm$0&00130    
&  1&02235&$\!\!\!\!\!\pm$0&00731   \\
$  \Sigma^*(-2;4 m_{\pi}^2) $
&  0&082682&$\!\!\!\!\!\pm$0&000660
&  0&068218&$\!\!\!\!\!\pm$0&000375
&  0&067882&$\!\!\!\!\!\pm$0&000106  
&  0&082346&$\!\!\!\!\!\pm$0&000553 \\
$ \Sigma^*(-3;4 m_{\pi}^2) $
&  0&008168&$\!\!\!\!\!\pm$0&000064
&  0&006773&$\!\!\!\!\!\pm$0&000037
&  0&006740&$\!\!\!\!\!\pm$0&000011    
&  0&008135&$\!\!\!\!\!\pm$0&000053 \\
\noalign{\smallskip}\hline
$ \Sigma(-1;4 m_{\pi}^2) $
&  1&03049&$\!\!\!\!\!\pm$0&00847   
&  0&83421&$\!\!\!\!\!\pm$0&00502
&  0&82978&$\!\!\!\!\!\pm$0&00127  
&  1&02650&$\!\!\!\!\!\pm$0&00716   \\
$ \Sigma(-2;4 m_{\pi}^2) $
&  0&082943&$\!\!\!\!\!\pm$0&000621   
&  0&068220&$\!\!\!\!\!\pm$0&000366
&  0&067866&$\!\!\!\!\!\pm$0&000106
&  0&082552&$\!\!\!\!\!\pm$0&000553\\
$ \Sigma(-3;4 m_{\pi}^2) $
&  0&008193&$\!\!\!\!\!\pm$0&000061 
&  0&006773&$\!\!\!\!\!\pm$0&000037 
&  0&006744&$\!\!\!\!\!\pm$0&000010  
&  0&008161&$\!\!\!\!\!\pm$0&000064 \\
\noalign{\smallskip}\hline
$ \Sigma^{\approx}(-1;4 m_{\pi}^2) $
&  1&02919&$\!\!\!\!\!\pm$0&00844
&  0&83292&$\!\!\!\!\!\pm$0&00499    
&  0&82833&$\!\!\!\!\!\pm$0&00127
&  1&02505&$\!\!\!\!\!\pm$0&00730   \\
$ \Sigma^{\approx}(-2;4 m_{\pi}^2) $
&  0&08269&$\!\!\!\!\!\pm$0&00062
&  0&06797&$\!\!\!\!\!\pm$0&00036  
&  0&06760&$\!\!\!\!\!\pm$0&00011
&  0&08229&$\!\!\!\!\!\pm$0&00059 \\
$ \Sigma^{\approx}(-3;4 m_{\pi}^2) $
&  0&00814&$\!\!\!\!\!\pm$0&00006
&  0&00672&$\!\!\!\!\!\pm$0&00004  
&  0&00668&$\!\!\!\!\!\pm$0&00001
&  0&00810&$\!\!\!\!\!\pm$0&00006 \\
\noalign{\smallskip}\hline
\end{tabular}
}
\end{table}

Finally, we are able to get the ``lattice extrinsic''
$\tilde{\cM}(-n)$ in terms of quantities accessible with the
lattice. The required relations read~\cite{deRafael:2014gxa}:
\ba
\tilde{\cM}(-1) &=& \Sigma(-1;s_0)
-\ln \frac{s_0}{m_\mu^2}\,\cM(-1)- \frac{m_\mu^2}{s_0}\,\cM(0)
+\frac{s_0}{m_\mu^2}\,\cM(-2) + \cdots
\crn
\tilde{\cM}(-2)  &=& -\Sigma(-2;s_0)
-\ln \frac{s_0}{m_\mu^2}\,\cM(-2)- \frac{m_\mu^2}{s_0}\,\cM(-1)+\frac12\,\left(\frac{m_\mu^2}{s_0}\right)^2\,\cM(0)
\crn && +\frac{s_0}{m_\mu^2}\,\cM(-3) + \cdots
\crn
\tilde{\cM}(-3) &=& \Sigma(-3;s_0)
-\ln \frac{s_0}{m_\mu^2}\,\cM(-3)- \frac{m_\mu^2}{s_0}\,\cM(-2)
+\frac12\,\left(\frac{m_\mu^2}{s_0}\right)^2\,\cM(-1)\crn && -\frac13\,\left(\frac{m_\mu^2}{s_0}\right)^3\,\cM(0)
+ \frac{s_0}{m_\mu^2}\,\cM(-4) +\cdots
\label{tildeMpred}
\ea
Adopting the Euclidean ``thresholds'' (\ref{Euclideanthreshold}), for
$s_0=m_{\pi^0}^2$ and $s_0=4 m_\pi^2$, the results up to 3$^{\rm rd}$ order
are given in Table~\ref{tab:sig}. These moments again are directly
accessible in lattice QCD and can provide important crosschecks. Again
contributions from regions above about 1~GeV are significant for
getting reliable estimates. 

It is interesting to note that the choice
$s_0=m_\mu^2$ leads to a simplification of the formulas
(\ref{tildeMpred}), particularly the second term with the log is then
absent. On the other hand for $s_0 > m_\mu^2$ the suppression factors
$m_\mu^2/s_0$ for the leading moments $\cM(0)$, $\cM(-1)$ etc. are
very welcome in reducing the largest cancellations. Note that the
$\tM(-n)$'s are an order of magnitude smaller than the corresponding
$\cM(-n)$'s for a given $n$ (see Table~\ref{tab:mom}). 

One observes a
strong dependence of the Euclidean integrals $\Sigma(-n; s_0)$ on the
infrared cutoff $s_0=Q_0^2$, reflected by the factor of about 5 between
the results for $s_0=m_{\pi^0}^2$ and $s_0=4
m_\pi^2$. In contrast,  the timelike
integrals, $\cM(-n)$ for example,  differ little once the $\rho$
peak in included in the integration range.  Changing $s_0$ from $4
m_\pi^2$ to $m_{\pi^0}^2$ increases the integral by the $\pi^0\gamma$
contribution, by about 0.5\% in $\cM(0)$. Typically, the dominant
$\rho$ resonance contribution in $R(s)$ in the Adler function $D(Q^2)$
appears completely smeared out leading to a steep monotonic increase
at low $Q^2$. This high sensitivity on $Q_0^2$, corresponding to
$x_{\rm min}$ in the integral (\ref{ADI}) where the limit $x_{\rm
min}\to 0$ is required and where the finiteness of the Adler function slope comes
into play. This high sensitivity is one reason why lattice QCD
calculations of $\amuh$ are so difficult in reducing uncertainties of
the needed extrapolations.
\begin{table}
\centering
\caption{Comparison of the moments $\tilde{\cM}(-n)$ as obtained
directly via (\ref{tildeMdire}) and via the moments expansion
(\ref{tildeMpred}) for $s_0=m^2_{\pi^0}$ and $s_0=4 m^2_{\pi}$ (in
units $10^{-5}$). Again the $\tilde{\cM}^*(-n)$ values are obtained
from the timelike integrals over $R(s)$, the other two from moments
which are accessible in lattice QCD: the $\Sigma(-n;s_0 )$ and $\cM(-n)$.}
\label{tab:tildeM}
{\scriptsize
\begin{tabular}{lr@{.}lr@{.}l|r@{.}lr@{.}l|r@{.}lr@{.}l|r@{.}lr@{.}l}
\noalign{\smallskip}\hline\noalign{\smallskip}
moment &
\multicolumn{4}{c}{data direct} &
\multicolumn{4}{c}{data HLS channels} &
\multicolumn{4}{c}{HLS model} &
\multicolumn{4}{c}{HLS + remainder}  \\
\noalign{\smallskip}\hline\noalign{\smallskip}
$\tilde{\cM}^*(-1)$
&-0&82592&$\!\!\!\!\!\pm$0&00516
&-0&79611&$\!\!\!\!\!\pm$0&00501
&-0&80054&$\!\!\!\!\!\pm$0&00113
&-0&83035&$\!\!\!\!\!\pm$0&00167 \\
$\tilde{\cM}^*(-2)$
&-0&026808&$\!\!\!\!\!\pm$0&000294
&-0&026644&$\!\!\!\!\!\pm$0&000294
&-0&027338&$\!\!\!\!\!\pm$0&000035
&-0&027502&$\!\!\!\!\!\pm$0&000035 \\
$\tilde{\cM}^*(-3)$
&-0&0013160&$\!\!\!\!\!\pm$0&0000228
&-0&0013149&$\!\!\!\!\!\pm$0&0000228
&-0&0013847&$\!\!\!\!\!\pm$0&0000017
&-0&0013858&$\!\!\!\!\!\pm$0&0000017   \\
\noalign{\smallskip}\hline\noalign{\smallskip}
$\left.\tilde{\cM}(-1)\right|_{m_{\pi^0}^2}$
&-0&84123&$\!\!\!\!\!\pm$0&00455
&-0&79592&$\!\!\!\!\!\pm$0&00535
&-0&80030&$\!\!\!\!\!\pm$0&00110
&-0&84561&$\!\!\!\!\!\pm$0&00146 \\
$\left.\tilde{\cM}(-2)\right|_{m_{\pi^0}^2}$
&-0&02222&$\!\!\!\!\!\pm$0&00054
&-0&02678&$\!\!\!\!\!\pm$0&00034
&-0&02749&$\!\!\!\!\!\pm$0&00005
&-0&02293&$\!\!\!\!\!\pm$0&00008 \\
$\left.\tilde{\cM}(-3)\right|_{m_{\pi^0}^2}$
&-0&003151&$\!\!\!\!\!\pm$0&000061
&-0&001222&$\!\!\!\!\!\pm$0&000007
&-0&001279&$\!\!\!\!\!\pm$0&000004
&-0&003207&$\!\!\!\!\!\pm$0&000011   \\ 
\noalign{\smallskip}\hline\noalign{\smallskip}
$\left.\tilde{\cM}(-1)\right|_{4 m_{\pi}^2}$
&-0&82828&$\!\!\!\!\!\pm$0&00508
&-0&79484&$\!\!\!\!\!\pm$0&00543
&-0&79916&$\!\!\!\!\!\pm$0&00110
&-0&83260&$\!\!\!\!\!\pm$0&00153  \\   
$\left.\tilde{\cM}(-2)\right|_{4 m_{\pi}^2}$
&-0&026801&$\!\!\!\!\!\pm$0&000339
&-0&026896&$\!\!\!\!\!\pm$0&000309
&-0&027616&$\!\!\!\!\!\pm$0&000035
&-0&027521&$\!\!\!\!\!\pm$0&000037 \\
$\left.\tilde{\cM}(-3)\right|_{4 m_{\pi}^2}$
&-0&001284&$\!\!\!\!\!\pm$0&000019
&-0&001258&$\!\!\!\!\!\pm$0&000021
&-0&001324&$\!\!\!\!\!\pm$0&000001
&-0&001350&$\!\!\!\!\!\pm$0&000001   \\
\noalign{\smallskip}\hline
\end{tabular}
}
\end{table}
The moments $\tilde{\cM}(-n)$'s ideally are not dependent on $s_0$, but
depend on $s_0$ though truncation errors in the moment expansion. Differences
as seen in Table~\ref{tab:tildeM} are not really small, but in view of
the strong $s_0$-dependence of the $\Sigma(-n;s_0)$ moments, they are quite
acceptable. Table~\ref{tab:cancellation} illustrates the composition
of the $\tM(-n)$ predictions in terms of Euclidean objects.

\begin{table}
\centering
\caption{Illustrating cancellations of moments according to
(\ref{tildeMpred}) (in units $10^{-5}$). The first term is
$\Sigma(-n,s_0)$ the second the log term and the last the truncated
$\cR(-n,s_0)$ from (\ref{Rapprox}) including moments up to $n=4$. The
second last term, which is proportional to $\cM(0)$ in any case
overcompensates the $\Sigma(-n,s_0)$ term. Note that uncertainties of
the leading terms are easily bigger than some of the subleading
contributions.}
\label{tab:cancellation}
{\scriptsize
\begin{tabular}{c|r|r|r||r|r|r}
\hline\noalign{\smallskip}
 & \multicolumn{3}{|c|}{$s_0=m^2_{\pi^0}$}
 &\multicolumn{3}{|c}{$s_0=4\,m^2_{\pi}$} \\
$n=$&\multicolumn{1}{c}{$1$} &\multicolumn{1}{c}{$2$} &\multicolumn{1}{c}{$3$} & 
\multicolumn{1}{c}{$1$} &\multicolumn{1}{c}{$2$} &\multicolumn{1}{c}{$3$} \\
\noalign{\smallskip}\hline\noalign{\smallskip}
$\Sigma(-n,s_0)$
&    5.449750  &    -1.771050  &     0.732931   &    1.022350  &    -0.082346  &     0.008135\\
$-\ln \frac{s_0}{m_\mu^2}\,\cM(-n)$
&   -0.116384  &    -0.004412  &    -0.000252   &   -0.461658  &    -0.017501  &    -0.001000\\
$\propto \cM(-2)$ &\multicolumn{1}{c|}{-}              &\multicolumn{1}{c|}{-}               &    -0.005519   &              &  &    -0.001290\\
$\propto \cM(-1)$ &\multicolumn{1}{c|}{-}              &    -0.145585  &     0.044602   &\multicolumn{1}{c|}{-}              &    -0.034042  &     0.002439\\
$\propto \cM(0)$&   -6.193043  &     1.897331  &    -0.775034   &   -1.448101  &     0.103737  &    -0.009908\\
$\cR(-n,s_0)$
&    0.014071  &     0.000788  &     0.000065   &    0.054808  &     0.002630  &     0.000276\\
\noalign{\smallskip}\hline\noalign{\smallskip}
$\tM(-n)$
&   -0.845606  &    -0.022928  &    -0.003207   &   -0.832601  &    -0.027521  &    -0.001350\\
\noalign{\smallskip}\hline
\end{tabular}
}
\end{table}

\section{How to get the moments $\tM(-n)$ directly from lattice QCD data?}
\label{sec:LQCDhowto}
We note that the moments $\tM(-n)$ are much smaller than the auxiliary
moments $\Sigma(-n; s_0)$. There are obviously large cancellations,
especially for small $s_0$, which enter as inverse power
weight factors of the moments $\cM(-n)$ to be subtracted from the
$\Sigma(-n; s_0)$. In fact these cancellations can be avoided to a
large extent.  How do we get the moments (\ref{sigma}) in terms of
lattice data?  A direct evaluation of (\ref{sigma}) in terms of the
Euclidean configuration space correlator, like (\ref{latticeTayCoe})
for the Taylor coefficients, is not available in this case. However,
if we assume that $\Pi(Q^2)$ has been determined by a Fourier
transform of the Euclidean correlator measured in configuration space,
in principle, (\ref{sigma}) can be integrated directly without problem
when we choose a finite infrared cutoff $Q_0^2>0$. Once we are given
$\Pi(Q^2)$ itself or as a Taylor series we may proceed as follows:
in order to get $\tM(-n)$ for a given $n$, subtract its Taylor expansion 
to order $n-1$ from $\Pi(Q^2)$, which defines
\be
\Pi(Q^2)^{\rm trunc}_n\equiv \Pi(Q^2)-
\sum_{j=0}^{n-1}\,\left(Q^2\right)^{(j+1)}\,\Pi_{j+1}\approx
\sum_{j=n}^{N}\,\left(Q^2\right)^{(j+1)}\,\Pi_{j+1}\,. 
\label{HVPtrunc}
\ee
As indicated, this also can be done if $\Pi(Q^2)$ is given as a Taylor
series to some order $N>n+1$.  In the latter case
\be
\Pi(Q^2)^{\rm trunc}_n/(Q^2)^{(n+1)}\approx
\sum_{j=0}^{N-n}\,\left(Q^2\right)^{(j)}\,\Pi_{n+j+1}\,, 
\label{Pitrunc}
\ee
can be Pad\'e improved. We then first evaluate the Taylor
moments to order $n-1$, which provides an expression in terms of
normal Taylor moments $\cM(-n)$ plus an integral over the subtracted
HVP function which, if given as a power series, has to be represented
in terms of appropriate Pad\'e approximants in order for the integral to
converge at large momenta. We thus obtain:
\ba
\Sigma^{\rm trunc}_n(-n;Q_0^2)
&=&-\sum_{j=0}^{n-1}\,(-1)^{(j+1)}\,\cM(-j)\,\frac{1}{n-j}\left(Q_0^2/m_\mu^2\right)^{j-n}\crn&&
+ \int\limits_{Q_0^2}^{\infty}\,\D Q^2\,\left(\frac{m_\mu^2}{Q^2}\right)^{n+1}\:
\left(-\frac{\Pi(Q^2)^{\rm trunc}_n}{Q^2}\right)\epo
\label{sigmatruncated}
\ea
If we subtract one more term for $j=n$ we obtain an UV divergent
result or, if regulated with a cutoff $Q_1^2$, we get a term
\be
\Sigma^{\rm trunc}_{n}(-n;Q_0^2)=
(-1)^{n+1}\,\cM(-n)\,\left(\ln Q_0^2/m_\mu^2-\ln Q_1^2/m_\mu^2\right)+
\Sigma^{\rm trunc}_{n+1}(-n;Q_0^2)\epo
\label{sigmatruncsing}
\ee
We thus reproduce the terms involving the moments $\cM(-n)$ in
(\ref{sigmafrommoms}), which means that evaluating the subtracted HVP function
(\ref{HVPtrunc}) is actually what essentially yields the moments
$\tM(-n)$, namely,
\be
\Sigma^{\rm trunc}_{n}(-n,s_0)=(-1)^{n+1}\,\left(\tM(-n)+ \ln
\frac{s_0}{m_\mu^2}\,\cM(-n) -\cR(-n,s_0)\right)\,.
\ee
Working out additional terms in the Taylor expansion may
be as effective and easier. Here we should remember that the remainders
$\cR(-n,s_0)$ can be made negligible by choosing $s_0$ small enough, as
they vanish in the limit $s_0 \to 0$. So it turns out that 
the moments $\tM(-n)$, by definition $s_0$--independent, are given by
the ``finite part'':
\ba
\tM(-n)&=&\lim_{s_0 \to 0}\,\left\{ 
(-1)^{n+1} \int\limits_{s_0}^{\infty}\,\D Q^2\,\left(\frac{m_\mu^2}{Q^2}\right)^{n+1}\:
\left(-\frac{\Pi(Q^2)^{\rm trunc}_n}{Q^2}\right)- \ln \frac{s_0}{m_\mu^2}\,\cM(-n) \right\}\epo
\label{tildeMlattice}
\ea
This is our master formula for the evaluation of the $\tM(-n)$ moments
in lattice QCD in particular. In this representation the only purpose
of the IR regulator $s_0$, to be chosen positive infinitesimal, is
to parametrize the logarithmic singularity. The latter is persisting since the $1/Q^2$ term 
at low $Q$ has been kept in the truncated HVP function, and remains
unaffected by going to the representation by suitable Pad\'e approximants. The latter have to
be chosen to behave at large $Q^2$ in accord with pQCD such that the
potential logarithmic UV singularity in (\ref{sigmatruncsing}) will be
absent.

What it simply amounts to is the following: let $\dalh (-Q^2)=-\Pi(Q^2)=
\sum_{j=0}^N p_j\,x^{j+1}$ 
with $x \equiv Q^2/m_\mu^2$ and $p_j= (-1)^j\,\cM(-j)$ from
(\ref{TaylorII}). Then we have to consider terms as listed in the
following tabular:
\begin{center}
\begin{tabular}{ccc}
\hline
moment & Taylor polynomial & term to be subtracted\\
\hline\noalign{\smallskip}
$\tM(-1)$ & $p_1+p_2\,x+p_3\,x^2\cdots$ & $\ln(x_0)\,\cM(-1)$\\ 
$\tM(-2)$ & $p_2+p_3\,x+p_4\,x^2\cdots$ & $\ln(x_0)\,\cM(-2)$\\ 
$\tM(-3)$ & $p_3+p_4\,x+p_5\,x^2\cdots$ & $\ln(x_0)\,\cM(-3)$\\ 
\multicolumn{2}{c}{$\cdots$}~~~~.
\end{tabular}
\end{center}
Then take the [1,2] Pad\'e for example of the polynomial in the list and integrate
$(-1)^{n+1}\,[1,2](x)/x$ over $x$ from $x_0=s_0/m_\mu^2$ to infinity and
subtract the IR sensitive term $\ln(x_0)\,\cM(-n)$ in order to get
$\tM(-n)$. For the [1,1] and [2,2] Pad\'es,  the integral obviously diverges.

This is our main results: the Mellin-Barnes moment expansion can be
implemented in a surprisingly straightforward manner by the evaluation
of (\ref{basicmoments}) or (\ref{latticeTayCoe}) for the Taylor
moments $\cM(-n)$ and by the evaluation of (\ref{tildeMlattice}) for
the log suppressed moments $\tM(-n)$.  A detour via the moments
$\Sigma(-n,s_0)$ and the remainders $\cR(-n,s_0)$ at the end turns out
to be superfluous, but may serve for crosschecks. 

\begin{figure}[t]
\centering
\includegraphics[height=6.6cm]{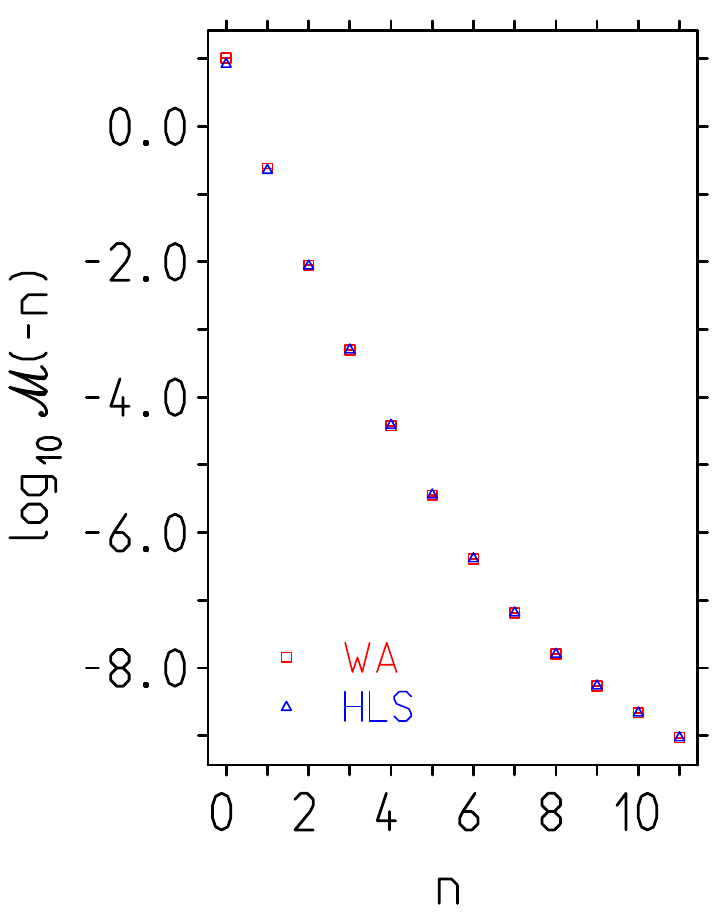}
\caption{The $log_{10}\,\cM(-n)$ are shown for $n=0,\cdots,11$. The HLS
model moments for the larger $n$ values  agree well with
the $\epm$-data estimates,
within uncertainties. Error bars are barely visible on top of the marks.}
\label{fig:moments12} 
\end{figure}

\begin{figure}
\centering
\includegraphics[width=0.5\textwidth]{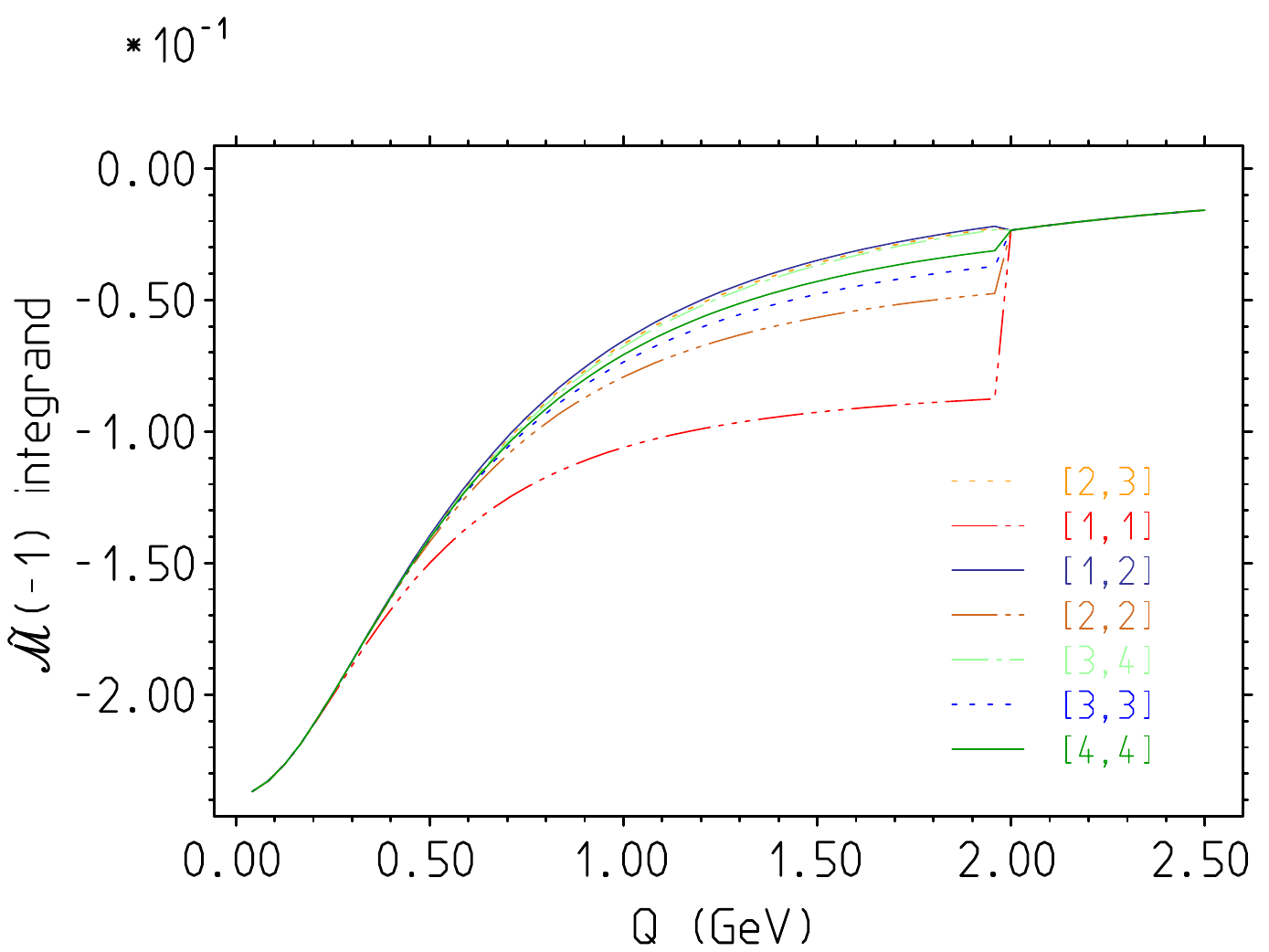}
\includegraphics[width=0.5\textwidth]{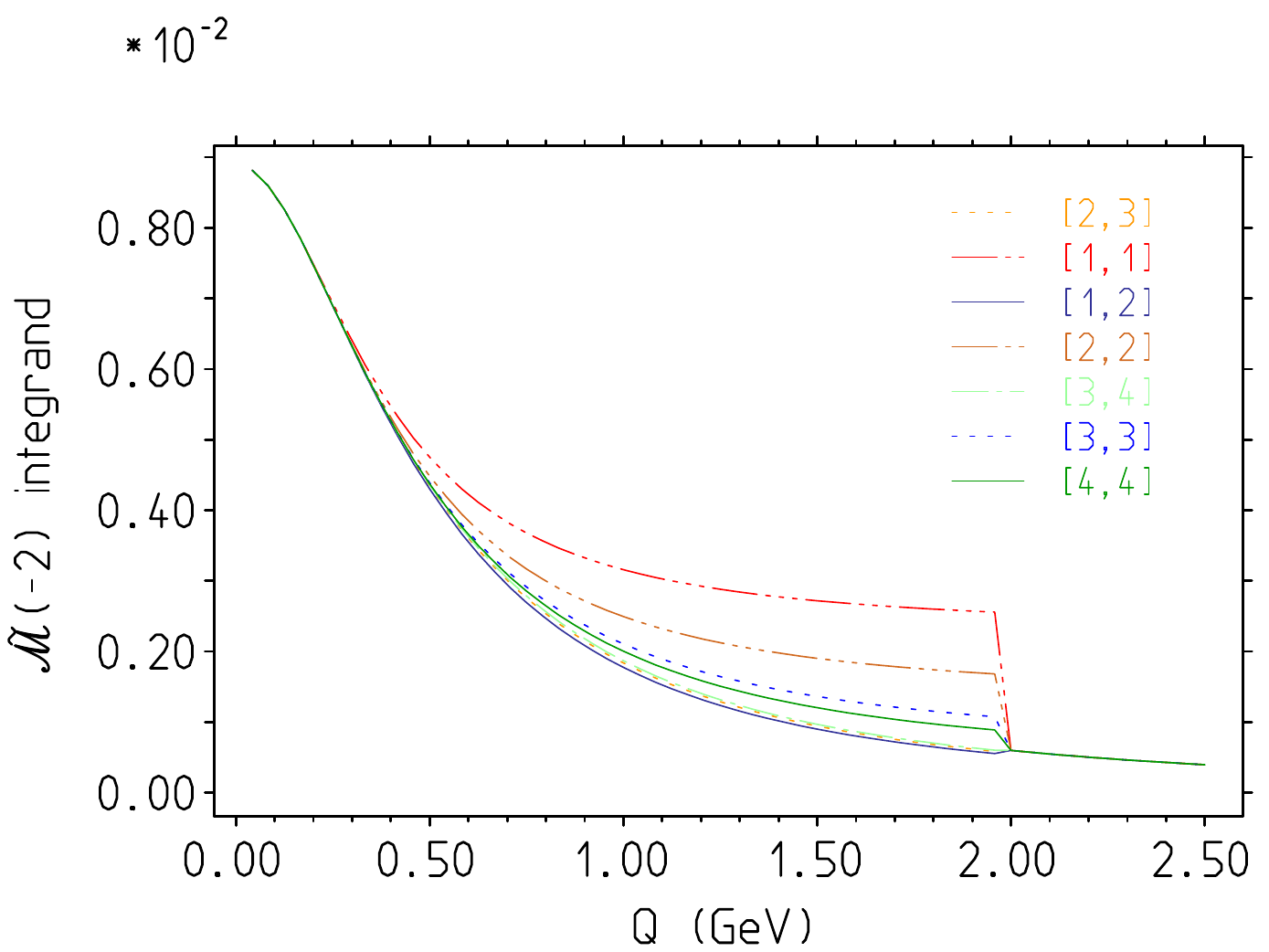}
\includegraphics[width=0.5\textwidth]{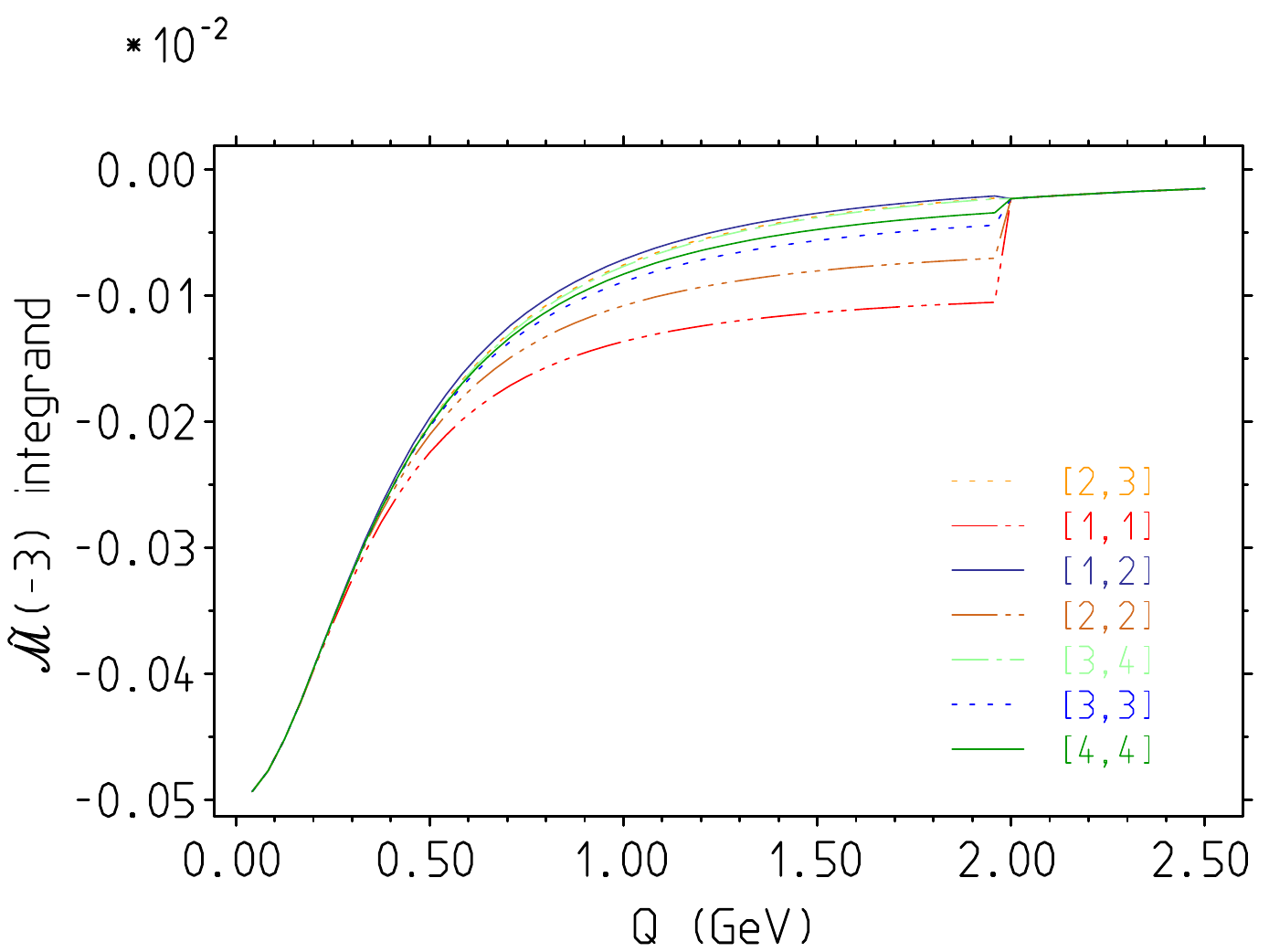}
\caption{The integrands of Eq.~(\ref{tildeMlattice}) with an extra
weight factor $x=Q^2/m_\mu^2$ for $\tM(-n)$ ($n=1,2,3$) as a
function of $Q$ and with a Pad\'e cutoff at 2~GeV. One can see that
the Pad\'e pairs [n-1,n]+[n,n] ($n=1,2,3,4$) nicely converge. }
\label{fig:tildeMint} 
\end{figure}

We have tested the truncated HVP approach by using $\dalh(-Q^2)$ based
on a world average (WA) compilation of the $\epm$ data (as available from
~\cite{hvpfunction}) together with the related Taylor 
moments\footnote{For the Taylor coefficients for the WA
compilation and the HLS model prediction we find\\[-3mm] 
\footnotesize
\begin{center}
\begin{tabular}{rlcl|lcl} 
\hline
$n$ & \multicolumn{3}{c}{WA compilation} & \multicolumn{3}{|c}{HLS model}\\
\hline
 $ 0$&$ 1.01962131E+01$&$\pm$&$ 6.693577E-02$&  $ 8.60436543E+00$&$\pm$&$ 1.303549E-02$\\
 $ 1$&$ 2.38190432E-01$&$\pm$&$ 1.257508E-03$&  $ 2.31974285E-01$&$\pm$&$ 3.137495E-04$\\
 $ 2$&$ 8.89142868E-03$&$\pm$&$ 5.749533E-05$&  $ 8.97346405E-03$&$\pm$&$ 1.138840E-05$\\
 $ 3$&$ 4.99117005E-04$&$\pm$&$ 4.018536E-06$&  $ 5.14676918E-04$&$\pm$&$ 6.561512E-07$\\
 $ 4$&$ 3.78809709E-05$&$\pm$&$ 3.333508E-07$&  $ 3.95581883E-05$&$\pm$&$ 5.245059E-08$\\
 $ 5$&$ 3.53345581E-06$&$\pm$&$ 2.971228E-08$&  $ 3.70101838E-06$&$\pm$&$ 6.216865E-09$\\
 $ 6$&$ 4.06928851E-07$&$\pm$&$ 2.867316E-09$&  $ 4.23990709E-07$&$\pm$&$ 1.378845E-09$\\
 $ 7$&$ 6.51188814E-08$&$\pm$&$ 4.102998E-10$&  $ 6.72641010E-08$&$\pm$&$ 4.547703E-10$\\
 $ 8$&$ 1.59771291E-08$&$\pm$&$ 1.260996E-10$&  $ 1.64295040E-08$&$\pm$&$ 1.722887E-10$\\
 $ 9$&$ 5.43416738E-09$&$\pm$&$ 5.033454E-11$&  $ 5.58818049E-09$&$\pm$&$ 6.967189E-11$\\
 $10$&$ 2.17223798E-09$&$\pm$&$ 2.125434E-11$&  $ 2.23581201E-09$&$\pm$&$ 2.941683E-11$\\
 $11$&$ 9.33132065E-10$&$\pm$&$ 9.276033E-12$&  $ 9.60971183E-10$&$\pm$&$ 1.283852E-11$
\end{tabular}
\end{center}

We refer to the comments at the end of Sect.~\ref{sec:BHLSmoments} for
what concerns the difference in the evaluations of the two sets of
moments which ideally should agree for the higher moments.  } to
order $n=12$, which allows us to calculate Pad\'e approximants [n-1,n]
and [n,n] up to $n=4$. In Fig.~\ref{fig:moments12} we display the
logarithm of the moments used in our analysis.
The Pad\'eized Taylor polynomials multiplied
by $x$ (to get ride of the $1/x$ singularity) are displayed in
Fig.~\ref{fig:tildeMint} and show a nice convergence. For the [n,n]
Pad\'es we need a high energy cutoff, above which we use the corresponding
truncated HVP function $\dalh(-Q^2)$. The [n-1,n] Pad\'es alone can be
integrated without a high energy cutoff. 
In any case, we
think our tables shed light on how the MBM approach works in Euclidean
space, where it is by far not as straightforward as in the timelike
domain. In case $\Pi(Q^2)$ is given as a Taylor expansion, at the end
it turns out that also the moments $\tM(-n)$ are determined by the normal
Taylor coefficients, just the indices get shifted according to
(\ref{Pitrunc}). Interestingly, the MBMs $\cM(-n)$, up to
normalization, are directly given by the Taylor coefficients $\Pi_{n}$
and there is no integration to be performed and thus no need for a
Pad\'eization. In contrast, the log suppressed MBMs $\tM(-n)$ are
obtained by integrating the truncated HVP, and if the latter is given
as a Taylor series, a Pad\'e improvement is necessary. After all the
$\tM(-n)$'s can also be obtained alone in terms of moments which can
be evaluated in configuration space via (\ref{latticeTayCoe}).

\begin{figure}
\centering
\includegraphics[width=0.75\textwidth]{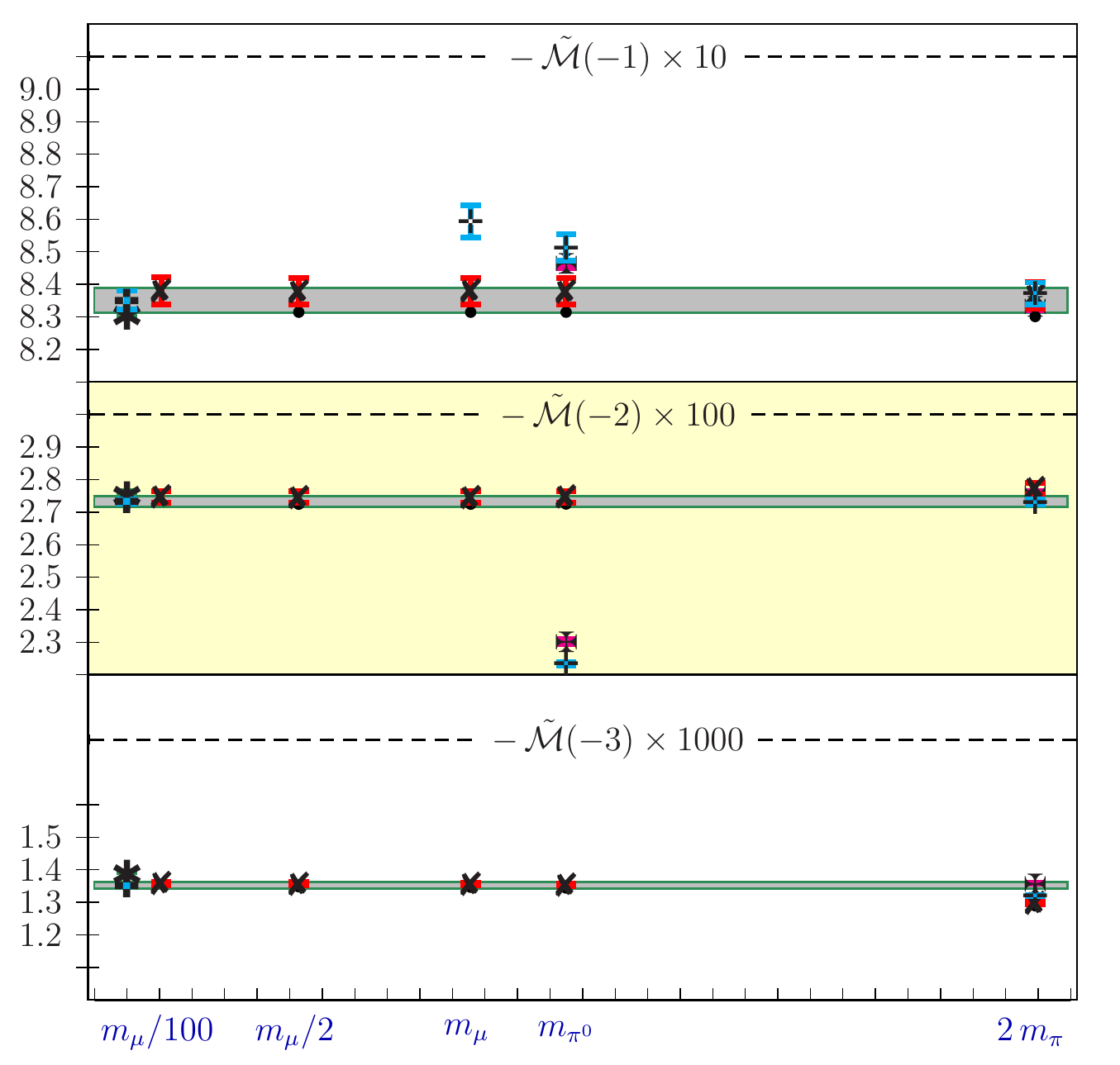}
\caption{The dependence of the moments $\tM(-n)$
evaluated via the truncated HVP on $\sqrt{s_0}$, together with the
direct (by definition $s_0$--independent) determination in terms of
data (\ding{60}), which also determines the error bands shown.  The
results obtained via (\ref{tildeMpred}) listed in
Table~\ref{tab:tildeM} are also displayed (marked by
\ding{64}, the $s_0$--independent ``HLS + remainder'' evaluation 
by \ding{81}). Values obtained via the Euclidean definition of
$\Sigma(-n; s_0)$ (marked by \ding{59}), within the ranges displayed,
could be evaluated reliably only for the largest $s_0$, where they
agree with the results collected in Table~\ref{tab:tildeM}. The
evaluation via the truncated HVP yield the results marked by a
$\bullet$ for the lower bound [3,4] Pad\'es. The upper bound Pad\'es
[n,n] n=1,2,3,4 lead to divergent integrals, if no cutoff is
applied. With a cutoff of 2~GeV the [3,4]+[4,4] averages, marked by a
\ding{56}, for $\tM(-1)$ moderately differ form the [3,4] Pad\'es
without cutoff, while for $\tM(-2)$ and $\tM(-3)$ the values agree
within uncertainties. The tilt in the $s_0$--dependence of the Pad\'e
estimates based on (\ref{tildeMlattice}), where the remainders
$\cR(-n;s_0)$ have been dropped as they vanish in the limit $s_0 \to
0$, completely disappears if one is including the remainders
$\cR(-n;s_0)$ (marks $\bullet$/\ding{56}, mostly not
distinguishable). For the closest point at $s_0=m_\mu^2/100$ the
remainder is negligible such that (\ref{tildeMlattice}) yields the
correct result for finite but small enough $s_0$. The lower order Pad\'e
pairs yield very similar results to the ones shown with larger errors, however.}
\label{fig:tildeMofs0} 
\end{figure}
As already mentioned earlier, by choosing $s_0=m_\mu^2$,
i.e. $x_0=1$, we get rid of the subtraction term! In this
representation there is no reason to choose $m_\mu^2/s_0$ small in
order to suppress the larger low--$n$ moments $\cM(-n)$, because the
subtraction is done on the level of the integrand in which case these
moments no longer appear. However, the remainders evaluated to be 
$\cR(-n,m_\mu^2)=8.6541(0.0555)\power{-3},4.8016(0.0314)\power{-4},3.7880(0.0330)\power{-5}$
for $n=1,2,3$ in units $10^{-5}$ are not negligible yet for $s_0=m_\mu^2$.

Note that uncertainties can barely be evaluated if we only are using
the lower bound [n-1,n] approximations, for which the integrals
converge. In order to get a handle to estimate the uncertainty we
cannot circumvent the consideration of the upper bound [n,n]
approximations as well. As the integrals of the latter do not
converge, we need to apply an UV cutoff, as we did when deriving the
results for $\amuh$ presented in Table~\ref{tab:TaylorPades}. We may
then take the mean and the deviation for pairs [n-1,n]+[n,n] and
should get reasonable error estimates for sufficiently large $n$,
besides a small additional contribution  from the high
energy tail, where we use $\dalh (-Q^2)$, as obtained from a world
average compilation of the $\epm$ data. The results obtained for
$\tM(-n)$ for $n=1,2,3$ based on the [3,4]+[4,4] Pad\'e approximants
are displayed in Fig.~\ref{fig:tildeMofs0}, together with the other
$\tM(-n)$ determinations presented before.

While the $\tM$'s can be neatly determined in terms of experimental
data or HLS predictions, the calculation of the $\tM$'s from purely
Euclidean LQCD data turns out to be problematic when we
attempt to use (\ref{tildeMpred}) together with (\ref{sigma}). One
problem is the choice of $s_0$, depending on the method applied, we
observe a substantial spread of the results. The most stable results
one obtains are for $s_0=4 m_\pi^2$. The methods based on the subtraction
of the moment expansion from the $\Sigma$ integrals are, in a way, 
complementary to the method based on integrating the integrand after
subtraction of the moment expansion (truncated HVP version). The first
works better for the larger $s_0$ values, but is of limited
reliability  because of the delicate cancellation pattern
illustrated in Table~\ref{tab:cancellation}. In contrast, for                  
sufficiently high Pad\'e approximants, the truncated HVP method works
perfectly and, for sufficiently low finite $s_0$, without the need to
know the reminders $\cR(-n,s_0)$.  

\begin{figure}
\centering
\includegraphics[height=8cm]{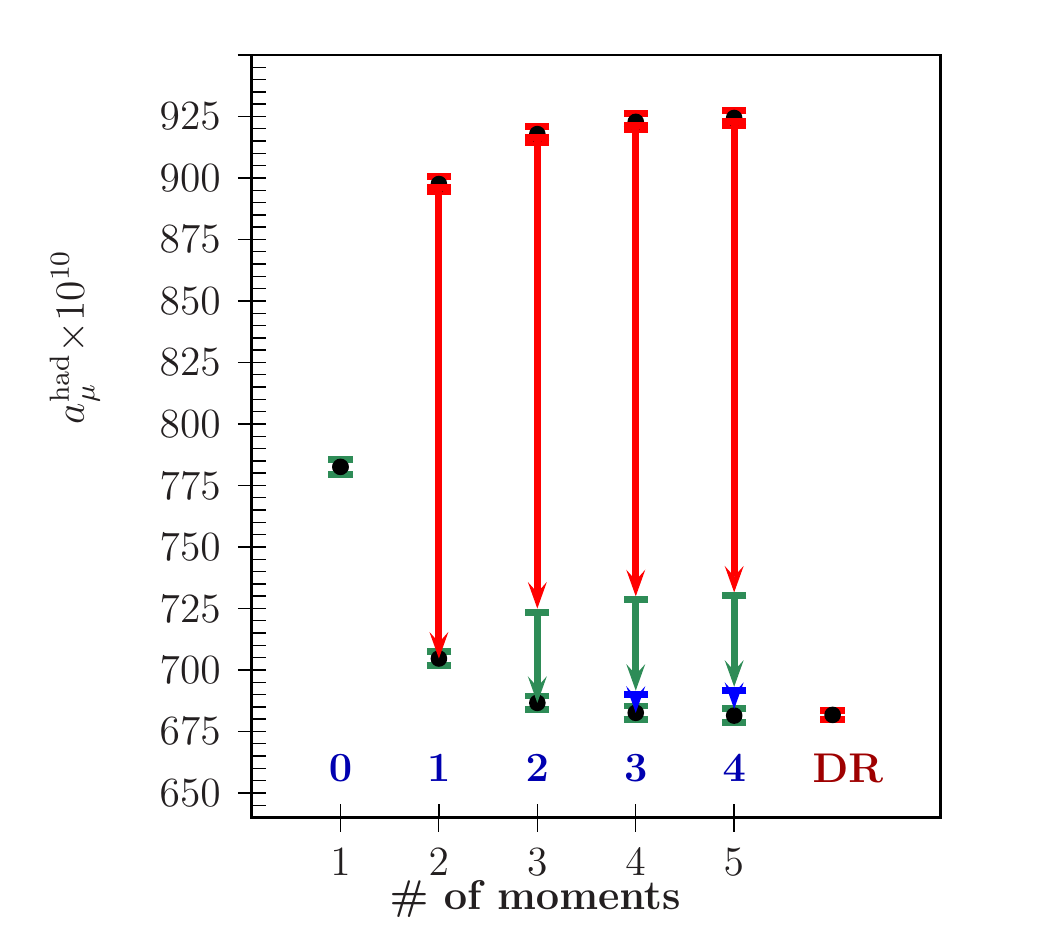}
\caption{This graph illustrates the relevance of the log suppressed
moments $\tM(-n)$ in calculating $\amuh$ via (\ref{tildeMint}). As we
know the starting point of the MBM expansion overestimates $\amuh$
substantially. All the normal (Taylor) moments $\cM(-n)$ are positive
and are corrections in the ``wrong direction''. Thus the $\tM(-n)$
moments are the ones not only to compensate the $\cM(-n)$
contributions, but also the ones which have to correct for the
overestimation we start with. This shows the importance to have
precise estimations of the $\tM(-n)$'s. The high points in the graph
are the ones given by the series (\ref{tildeMint}) where the $\tM$'s
are dropped, while the lowest ones represent the full result
(\ref{tildeMint}). The $\tM$'s are included consecutively for
$n=1,2,3$ (red, green, blue) from top to bottom. The $\tM(-4)$
contribution is too small to be displayable here.}
\label{fig:gm2moments1} 
\end{figure}

The Pad\'e approximants obtained from a low energy expansion in
general fail to be reliable at higher momentum transfer.  It looks
bold to predict what happens above the $\rho$ resonance from
information which zooms into what is happening below it, however, as
illustrated by Fig.~\ref{fig:tildeMint}, this is precisely what
seems to work\footnote{We quote {Numerical Recipes} commenting an
example with five terms of a power series in $x$: Why does this work? 
Are there not other functions with the same first five terms in their
power series, but completely different behavior in the range (say)
$2<x<10$? Indeed there are. Pad\'e approximation has the uncanny knack
of picking the function
\textit{you had in mind} from among all the
possibilities. \textit{Except when it doesn't}! That is the downside
of Pad\'e approximation: it is uncontrolled. There is in general no
way to tell how accurate it is, or how far out in $x$ it can usefully
be extended. It is a powerful, but in the end still mysterious,
technique~\cite{NR}.}. If we want to avoid the need of Pad\'e
approximations the method based on (\ref{sigmafrommoms}) solved for
$\tM$ together with the expansion (\ref{Rapprox}) is adequate. This
method requires an optimized choice of $s_0$, because moments enhanced
by powers of $m_\mu^2/s_0$ appear together with higher moments
weighted by factors $s_0/m_\mu^2$ in (\ref{tildeMpred}). For lower
$s_0$ values the cancellations in (\ref{tildeMpred}) grow and may
cause numerical problems. Fortunately, the moments $\Sigma(-n; s_0)$
directly evaluated via their definition (\ref{sigma}), agree rather
well with their estimations in terms of timelike data via
(\ref{Sigmadire}), as it should be. This can be concluded from the
numerical crosscheck presented in Table~\ref{tab:sigspacelike}.
\begin{table}
\centering
\caption{A numerical consistency test: Comparison of 
the timelike with the spacelike evaluations of the moments
$\Sigma(-n; s_0)$ and remainders $\cR(-n; s_0)$ in units
$10^{-5}$. Here we use moments obtained for the compilation used to
evaluate $\dalh(-Q^2)$ for $s_0=m_{\pi^0}^2$ in the upper part and for
$s_0=4 m_{\pi}^2$ in the lower part. By $\Sigma^*(-n; s_0) $ and
$\cR^*(-n; s_0) $ we denote the result from (\ref{Sigmadire}) and
(\ref{Rdire}), respectively. The version $\Sigma^\approx(-n; s_0)$ and
$\cR^\approx(-n; s_0)$ denote the LQCD appropriate evaluations of
(\ref{sigma}) and the truncated expansion (\ref{Rapprox}) including
moments up to $n=4$.  }
\label{tab:sigspacelike}
{\scriptsize
\begin{tabular}{lr@{.}lr@{.}l|r@{.}lr@{.}l||r@{.}lr@{.}l|r@{.}lr@{.}l}
\noalign{\smallskip}\hline\noalign{\smallskip}
moment &
\multicolumn{4}{c}{$\Sigma^*$ (\ref{Sigmadire})} &
\multicolumn{4}{c}{$\Sigma^\approx$ (\ref{sigma})} &
\multicolumn{4}{c}{$\cR^*$ (\ref{Rdire})} &
\multicolumn{4}{c}{$\cR^\approx$ (\ref{Rapprox})}  \\
\noalign{\smallskip}\hline\noalign{\smallskip}
$n=1 $
&5&5032&$\!\!\!\!\!\pm
$0&0380 
&5&4993&$\!\!\!\!\!\pm
$0&0366
&0&013895&$\!\!\!\!\!\pm
$0&000089
&0&013901&$\!\!\!\!\!\pm
$0&000089
\\
$\phantom{n=}2 $
&1&7882&$\!\!\!\!\!\pm
$0&0121 
&1&7869&$\!\!\!\!\!\pm
$0&0116
&0&00076861&$\!\!\!\!\!\pm
$0&00000615
&0&00076410&$\!\!\!\!\!\pm
$0&00000495
\\
$\phantom{n=}3 $
&0&7400&$\!\!\!\!\!\pm
$0&0050 
&0&73944&$\!\!\!\!\!\pm
$0&00479
&0&00005762&$\!\!\!\!\!\pm
$0&00000051
&0&000061822&$\!\!\!\!\!\pm
$0&000000539
\\
\noalign{\smallskip}\hline\noalign{\smallskip}
\noalign{\smallskip}\hline
$\phantom{n=}1 $
&1&0328&$\!\!\!\!\!\pm
$0&00780
&1&03228&$\!\!\!\!\!\pm
$0&00749
&0&052864&$\!\!\!\!\!\pm
$0&000337
&0&054193&$\!\!\!\!\!\pm
$0&000355
\\
$\phantom{n=}2 $
&0&083181&$\!\!\!\!\!\pm
$0&000603
&0&083108&$\!\!\!\!\!\pm
$0&000581
&0&0028220&$\!\!\!\!\!\pm
$0&0000222
&0&0025609&$\!\!\!\!\!\pm
$0&0000150
\\
$\phantom{n=}3 $
&0&0082175&$\!\!\!\!\!\pm
$0&0000588
&0&0082098&$\!\!\!\!\!\pm
$0&0000566
&0&00020577&$\!\!\!\!\!\pm
$0&00000182
&0&00026439&$\!\!\!\!\!\pm
$0&00000230
\\
\noalign{\smallskip}\hline
\end{tabular}
}
\end{table}
Concerning methods which require Pad\'e approximations to be used as a
tool for the extrapolation towards higher momenta we have to be aware
that the HVP ``modeling'' never is really good, because $\Pi(Q^2)$
grows logarithmically and not powerlike. The need of the truncated HVP
in the evaluation of the $\tM$'s also reveals, that one needs more
Taylor moments than we might have expected in order to reach a desired
precision\footnote{Considering the truncated HVP required for the
calculation of $\tM(-n)$ we loose $n$ Taylor coefficients and since we
need [n-1,n] and [n,n] type Pad\'e approximants, which requires $2n+1$
coefficients, a [4,4] Pad\'e for $\tM(-3)$ requires 12 Taylor
coefficients.}. We should be aware also of the fact that the MBM estimation
(\ref{tildeMint}) of $\amuh$ is very sensitive to the $\tM$'s as
illustrated in Fig.~\ref{fig:gm2moments1}. It reveals that the MBM
method has its weak points here.

Although the Mellin-Barnes moments approach shows excellent
convergence,  it looks to be quite elaborate as one has to determine
quite a number of moments in order to get reliable results. As
advocated in~\cite{Jegerlehner:2001wq,Jegerlehner:2008rs} (also see
~\cite{EJKV98} and references therein), the best and simplest check of
lattice QCD data is to compare the results with the Adler function as
it enters in the representation (\ref{ADI}) and as it actually has
been performed in~\cite{DellaMorte:2014rta}, recently. An up-to-date
evaluation of the ``experimental'' Adler function $D(Q^2)$ is
available via the link~\cite{Adlerfunction}. We remind that the Adler
function asymptotically tends to a constant at high $Q^2$, which means
that Pad\'e approximants applied to the Adler function in principle
can be properly matched to QCD asymptotics. This is another advantage
of working with $D(Q^2)$ rather than with $\Pi(Q^2)$ (compare
Figs.~\ref{fig:HVPPades} and \ref{fig:ADPPades} in this context). 

\section{BHLS  Evaluation of the I=1 Component of the HVP}
\label{sec:BHLSisovec}
Most LQCD calculations of the Euclidean HVP function attempt to
derive the isovector part as the leading contribution in a first
step. It is, therefore, desirable to have an ``experimental''
counterpart, which allows one to compare results. However, on the data
side, a separation of the I=1 part from the $\epm$ data as well as
its determination in terms of the $\tau$ data is not straightforward
(missing channels, electromagnetic effects). A corresponding
evaluation in the HLS model looks to be much more reliable and is
presented in this section.
\subsection{Reconstruction of HVP From Normal Fits \& $\tau$+PDG}
\indent \indent The model results referred to in the above Sections have been all derived
by running in the standard mode the broken HLS (BHLS)  model as defined in \cite{Benayoun:2011mm}
and recently improved in \cite{Benayoun:2015gxa}. This improvement deals with 
the need to properly account, in the fitting procedure,  for the 
special character of the overall normalization uncertainties strongly affecting
 the most recent $e^+ e^- \ra \pi^+\pi^-$  data samples. 
 This last study also provided a new update
of the BHLS analysis  \cite{Benayoun:2012wc} by including the
spectra recently published by KLOE \cite{KLOE12} and BESIII \cite{BESIII}.

The standard running mode of the BHLS fit procedure is a {\it global} fit which covers
simultaneously all the physics channels embodied within the BHLS model. As already noted, 
these represent the  six $e^+ e^-$ annihilation channels  to  
$\pi^+\pi^-$, $\pi^0\gamma$, $\eta \gamma$, $\pi^+\pi^-\pi^0$, $K^+K^-$ and $K_L K_S$,
the $\tau \ra \pi^\pm \pi^0 \nu$ decay and a few additional pieces of
light meson decay information \cite{Benayoun:2011mm}. The yielded fit quality  is high  
\cite{Benayoun:2015gxa} 
as reflected by the probability and the average  $\chi^2$ per data point for all
the physics channels addressed. 

For the present purpose, it is worth noting that the (BHLS) model description closely follows
the information which can be directly derived from the existing experimental data, as  well reflected 
by the two central data columns in Tables \ref{tab:mom} and \ref{tab:momres} and also
by Fig.~\ref{fig:moments12};  this indicates
 that the model dependence of the numerical results should be quite marginal.

The purpose of the present work is to relate phenomenology and the calculations which can be 
achieved within the framework of lattice QCD.  Our aim is twofold~: compare
LQCD predictions with data on the one hand, and  on the other hand,
initiate -- with BHLS -- examining the  relevance of 
effective models versus LQCD. For this purpose, 
 Figure 5 in \cite{Chakraborty:2016mwy} clearly indicates  that model predictions
 for $\amuh$  \cite{Jegerlehner:2015stw,Benayoun:2015gxa} derived using Effective 
 Lagrangians or elaborate  data handling (as \cite{Hagiwara:2011af}), are in fair agreement
with LQCD estimates; how this general agreement will evolve with the increasing
accuracy of lattice computations is an important issue to follow. 

Nevertheless, going closer to  what can be derived within the
lattice computation framework as it presently stands is certainly valuable.
For instance,  if one could motivatedly single out the isovector component of $\amuh$,
its comparison with LQCD predictions could be directly performed. In this prospect,
identifying  (and switching off) the IB effects at work in the experimental 
data and splitting up reliably the Isospin 0 and 1 components of $\amuh$ is
of particular relevance.  Obviously, such
a program can hardly be performed directly with the measured data, while it looks   
 in the realm of   effective models. As the BHLS model accounts well for 
 a large amount of experimental data in various physics channels, 
  such a procedure deserves to be attempted with it.

\subsection{Isospin Breaking and the $\tau$+PDG Approach}
\indent \indent 
It has been shown
\cite{Benayoun:2015gxa,Benayoun:2012wc,Benayoun:2015hzf} that the (dominant) contribution of
the $\pi^+\pi^-$ intermediate state to $\amuh$  can be well estimated\footnote{In the energy range
limited upward by 1.05~GeV, the domain of validity of the HLS model  \cite{Harada:2003jx}. This 
is not a real limitation to compare with  precise LQCD estimates.}  
without using the $e^+ e^- \ra \pi^+\pi^-$  experimental 
spectra. The pion form factor $F^\tau_\pi(s)$ in the $\tau \ra \pi^\pm \pi^0 \nu$ decay
can be almost exactly identified with  the hypothetical isospin symmetric pion form factor $F_\pi(s)$;
indeed, as far as the pion form factor is concerned, IB effects generated by the pion mass splitting are located
only in the pion loop entering the charged $\rho$ 
propagator\footnote{As also for the Kaon mass splitting in the Kaon loops.}. 

The issue solved by BHLS is to provide a {\it global} framework and a fitting tool 
able to derive the pion form factor $F^e_\pi(s)$
in the $e^+ e^-$ annihilation from fitting $F^\tau_\pi(s)$ and a few pieces of decay information carrying
the isospin breaking (IB) content at work in $e^+ e^- \ra \pi^+\pi^-$. 
These IB pieces of information 
are\footnote{ \label{phi_info}
Actually,  the listed  pieces of information for the $\phi$ meson
are used in  the standard running of the BHLS fitting code because
no published experimental dipion spectrum covering the $\phi$ mass region
is presently available.}~:  

\begin{itemize}
\item
{\bf (i)}  the $V \ra \pi^+\pi^-$ partial widths  for $V=\omg,~\phi$, 
\item
{\bf (ii)} the products  $\Gamma(V \ra \pi^+\pi^-) \times \Gamma(V \ra e^+ e^-)$  for $V=\omg,~\phi$, 
\item
{\bf (iii)} the $\rho^0 \ra e^+ e^-$ partial width,
\end{itemize}
\noindent which can be extracted from the Review of Particle Properties (RPP) \cite{RPP2012}.
Of course,  $F^\tau_\pi(s)$ depends on the  $[\pi^\pm \pi^0]$ (and $[K^\pm K^0]$)
loop(s), but the prediction for  $F^e_\pi(s)$ accounts automatically for its dependence
 upon the $[\pi^+ \pi^-]$ (and $[K^+ K^-]$,  $[K^0 \overline{K}^0]$) loop(s) with thresholds
 at their physical masses.

In principle, the $\rho^0 \ra e^+ e^-$ information is already contained inside 
the other channels encompassed within the BHLS framework and could be avoided; however, 
as this coupling  has a marginal impact in these other processes,  a more precise 
information is mandatory.

\begin{figure}
\centering
\caption{The pion form factor (PFF) data compared to the $\tau$+PDG prediction
and to the global fit. 
The upper sequence displays the $\tau$+PDG prediction  $F^e_\pi(s)$ for the
PFF  in $e^+ e^-$ annihilations together with the 
indicated data superimposed. The middle sequence displays the difference
between this prediction and the data. The lower sequence shows the (true) residual
plots, e.g. the difference between the global fit solution to  $F^e_\pi(s)$
and the data. Both kinds of residuals are corrected (see text).
 The average $\chi^2$ distances of the prediction and of the fit solution to the
data samples  are indicated in each panel.
\vspace{-3cm}}
\includegraphics[width=\linewidth]{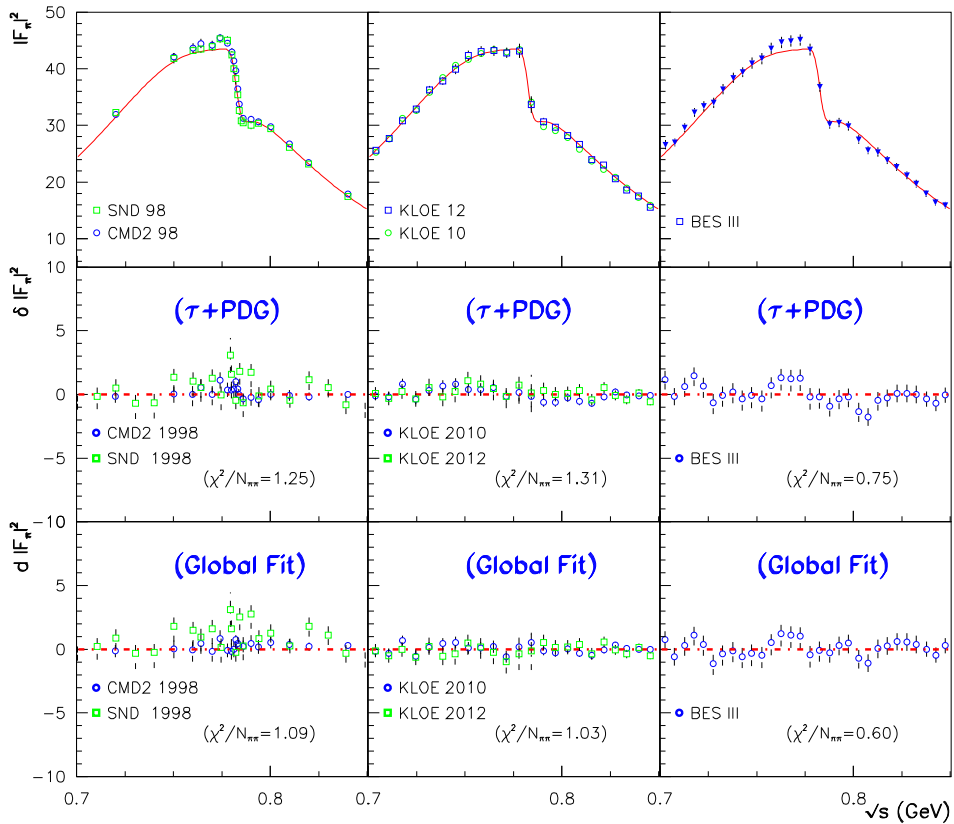}
\label{taupred} 
\end{figure}

The results which summarize the  $\tau$+PDG {\it prediction} for $F^e_\pi(s)$ are shown
in Fig.~\ref{taupred} and deserve some comments about how well IB effects accommodate
the BHLS framework. The upper three panels display the $\tau$+PDG 
predicted $F^e_\pi(s)$ function together with the $\pi^+ \pi^-$
data; one should keep in mind that all the $\pi^+ \pi^-$ experimental spectra are
 {\it excluded from the BHLS global fit} when running in the $\tau$+PDG mode. As noted in 
 \cite{Benayoun:2012wc,Benayoun:2015gxa}, the picture which arises from these plots
 is, at the observed level, surprisingly successful, showing that  the IB information  
 requested by BHLS is carried solely by the data pieces listed above in {\bf (i--iii)}.  This statement
 is enforced by   the middle sequence of panels in Fig.~\ref{taupred} where one has displayed 
 the plots of the difference between the experimental data
and the $\tau$+PDG prediction of the BHLS model; indeed, after the 
canonical\footnote{
\label{norm}
Without going into details and references which can be found in \cite{Benayoun:2015gxa}, let us sketch
how a global scale uncertainty  should be accommodated. For any function $f(s)$,
the unbiased residuals $\Delta f(s)$ are, in principle,  derived from the raw residuals 
$f_{exp}(s) - f_{fit}(s)$  via 
$\Delta f(s)= f_{exp}(s) - f_{fit}(s)-\lambda f_{true}(s)$, where $\lambda$ is
the  scale uncertainty which can be derived using fit results \cite{Benayoun:2012wc}.  
When $f_{true}(s)$ 
is unknown -- which is a rather common situation -- an iterative fit procedure has been
 shown to lead to a $f_{fit}(s)$ close enough to $ f_{true}(s)$ that 
$f_{exp}(s) - (1+\lambda) f_{fit}(s)$  is a very good approximation
of $\Delta f(s)$. } correction for the global scale uncertainties, 
the "pseudo-residual" distributions (they do not follow from a fit involving
the measured $\pi^+ \pi^-$ spectra)
  are shown quite satisfactorily spread around the 
zero level; this is especially striking for the KLOE or BESIII 
spectra\footnote{The decay information  {\bf (i--iii)} extracted
from the RPP \cite{RPP2012} is driven by the CMD2 and SND pion form factor
spectra \cite{CMD203,CMD206,SND06}  and totally independent of
 their analogs from KLOE or BESIII.
} which are statistically free of any correlation with any of the data samples or
decay information  running in the $\tau$+PDG  fit mode.

 In order to substantiate the quality of the prediction, the lowest sequence of
panels shows the corresponding (real) residual distributions\footnote{By real residuals,
we mean the normalized differences$^{\ref{norm}}$ between the $e^+ e^-$ data and the $F^e_\pi(s)$ 
derived from the global fit involving also the $e^+ e^- \ra \pi^+\pi^-$ data samples.}. 
These are as well
 centered around  the zero residual level as the pseudo-residuals. The improvement
 provided by the global fit compared to the $\tau$+PDG  fit mode is in the dispersal
 of the residuals, larger for the pseudo-residuals  than for the real residuals,
as evidenced by comparing the respective average $\chi^2$'s which can be read off the
various panels in
Figure \ref{taupred}. This improvement -- even if small --  is not really unexpected as 5  
(actually$^{\ref{phi_info}}$  3) 
IB pieces of  information are replaced by  $\simeq 320$ data points.

 Stated otherwise,
relying on the dipion spectra collected by ALEPH, CLEO and Belle and on some decay
data, one yields a precise determination of $F^e_\pi(s)$ and a good
estimation of $\amuh(\pi^+\pi^-)$, as shown in  \cite{Benayoun:2015gxa,Benayoun:2015hzf}. 
The $\tau$+PDG approach of BHLS provides results displayed in Table \ref{tab:gm2_hls}.

\begin{table}
\centering
\caption{The LO-HVP contribution in terms of moments in units of $10^{-10}$.
The rightmost pair of data columns are derived by dropping out the effects
generated by the Isospin breaking (IB) terms. The integration is performed
over the usual BHLS range, {\it i.e.} from $m_{\pi^0}$ to  1.05~GeV.
}
\label{tab:gm2_hls}
{\footnotesize
\begin{tabular}{lc|c|c|c|c}
\noalign{\smallskip}\hline\noalign{\smallskip}
\phantom{moment} &
{Exp. data } &\multicolumn{2}{c |}{HLS Fits} 
&\multicolumn{2}{c}{HLS Fits (IB terms removed)} \\
\phantom{moment} &
{HLS scope} &
{Standard Fit} &
{$\tau$+ PDG} &
{$\gamma$ \& I=0 \& I=1}  &
$\gamma$ \& $I=1$ \\
\noalign{\smallskip}\hline\noalign{\smallskip}
 $\amuh(0)$&  668.00$\pm$    3.83&  666.22$\pm$    1.01&  665.27$\pm$  1.70 & 631.95$\pm$  0.95 &  552.15$\pm$  0.75\\
 $\amuh(1)$&  594.11$\pm$    3.56&  592.50$\pm$    0.90&  591.72$\pm$  1.47 & 562.08$\pm$  0.84 &  489.66$\pm$  0.67\\
 $\amuh(2)$&  576.51$\pm$    3.41&  574.64$\pm$    0.88&  573.85$\pm$  1.44 & 545.08$\pm$  0.83 &  473.49$\pm$  0.65\\
 $\amuh(3)$&  572.65$\pm$    3.35&  570.58$\pm$    0.88&  569.86$\pm$  1.45 & 541.25$\pm$  0.82 &  469.70$\pm$  0.64\\
 $\amuh(4)$&  571.59$\pm$    3.33&  569.41$\pm$    0.87&  568.76$\pm$  1.42 & 540.16$\pm$  0.82 &  468.62$\pm$  0.64\\
\noalign{\smallskip}\hline\noalign{\smallskip}
  $\amuh$ &  570.68$\pm$    3.67&  568.95$\pm$     0.89 & 568.11$\pm$  1.45  &  539.56 $\pm$ 0.81 &  468.03$\pm$ 
  0.65\\
\noalign{\smallskip}\hline
\end{tabular}
}
\end{table}

The first data column in this  Table  reproduces the integration 
of the experimental data covered by BHLS  and the second data column shows
the results coming from integrating the solution to the standard (normal) fit; these have 
already been given in the two central columns of Table \ref{tab:momres} and are
reminded for convenience. The third data column in Table
\ref{tab:gm2_hls} displays the results derived by running the fit procedure
in the $\tau$+PDG  mode just sketched. One should note the closeness of the
corresponding numbers in the second data column (derived by fitting the rich
set of  $\pi^+\pi^-$ data samples from  the
CMD--2, SND, KLOE and BESIII Collaborations) and the third data column 
(relying on the $\tau$ data and some limited RPP information). The really new  information 
here is to remark that the statistics of the $\pi^+\pi^-$ data samples allows
to improve the uncertainty by $\simeq 50$\%,  while the central values shifts
by less than $1\sigma$. Another improvement is that the distributions
from which the central values for the $\amuh(n)$'s  and their standard deviations are extracted are much
closer to Gaussians in the standard mode than when running the $\tau$+ PDG 
mode\footnote{The distribution obtained by sampling 
the fit parameters on a multidimensional Gaussian with the fit covariance matrix are 
very close to 'perfect' Gaussians in the standard mode running, despite non--linearities; 
in the $\tau$+ PDG mode running,
small non--Gaussian tails distort the parameter distributions; referring to the last line of
Table \ref{tab:gm2_hls}, the numerical estimate of the mean value and r.m.s.
of the distribution gives $568.28 \pm  2.31$ instead of its Gaussian
fit result $568.11 \pm1.45$ .}.

\subsection{Inverting The $\tau$+ PDG Approach}
\indent \indent
The $\tau$+ PDG approach, which allows to plug IB effects within  the 
isovector part of the pion form factor  $F_\pi(s)(=F_\pi^\tau(s))$ 
and in the photon HVP ($\amuh$), provides also a way back
to restore Isospin symmetry  in the cross sections used  to evaluate the $\amuh(n)$'s and in
the MBM moments (or in the Taylor expansion series coefficients). 

As LQCD calculations generally focus on the I=1 part 
of the HVP and neglect IB effects,  one can expect BHLS to extract
from data quantities which can be the most directly compared with LQCD
predicted HVP values and moments. Since the higher moments ($n\ge 2$)
in the moment analysis are given by the HLS accessible contributions
within uncertainties (see Table~\ref{tab:mom}), the corresponding LQCD
analysis concerns just the range of validity of the BHLS model.
 
In order to construct the requested amplitudes, the fit parameter values and the
parameter error covariance matrix should be those of the global fit in the standard
mode running where all data are submitted to fit, as in \cite{Benayoun:2015gxa} for
instance. Indeed, the standard fit is supposed to provide the basic parameters
of the unbroken HLS model (like the parameters named $g$, $a$, the FKTUY parameters
\cite{FKTUY,Harada:2003jx} $c_i$, \ldots), beside the strictly speaking breaking parameters.

Then, restoring Isospin conservation is performed first by switching off the IB 
parameter\footnote{These are essentially the model parameters \cite{Benayoun:2011mm} named 
$\Delta_A$, $\Delta_V$. Taking the pion form factor in the $\tau$ decay 
as reference, implies to let $\Sigma_V$  vary  within its allowed range.} while keeping 
the others.
Isospin symmetry also imposes to cancel out the I=1 components inside the 
$\omg$ and $\phi$ mesons; this turns out to forbid the  $\omg/\phi \ra \pi^+\pi^-$ decays.  
 Within BHLS, these decays are generated via the difference between the charged and neutral Kaon 
 loops; then, as restoring  Isospin conservation implies to impose $m_{K^\pm}=m_{K^0}$, 
 this loop difference should be canceled out, preventing  the $\rho^0-\omg$ and
$\rho^0-\phi$ dynamical mixings; so, the  vanishing of the mixing angles 
\cite{Benayoun:2011mm}  $\alpha(s)$ and $\beta(s)$     
cancels out the $\omg/\phi \ra \pi^+\pi^-$ couplings;
however, the dynamical mixing in the  $\omg-\phi$ sector
still survives as it is driven by the sum of
the Kaon loops. 

\vspace{0.5cm}

At this stage, the BHLS Isospin conserved amplitudes contain   well defined I=0
(tagged by the  couplings to  either of the $\omg$ or $\phi$ mesons) and I=1
(tagged by the coupling to $\rho^0$) components;  the direct coupling of 
photons  to the hadronic final state also survives in our amplitudes.

The results for $\amuh(n)$ referring to the two configurations named resp.
$\gamma$+(I=0)+(I=1) and $\gamma$+(I=1) are  derived by using generated Monte Carlo data 
samples and the results are given in the last two data columns of Table 
\ref{tab:gm2_hls}. In this Table the integration is performed from $m_{\pi^0}$ to 1.05~GeV. 
Moreover, the correlations between the parameters considered and those which are canceled
out are accounted for at the Monte Carlo generation level. 

The last line in Table \ref{tab:gm2_hls} indicates that IB effects can be estimated
to $29.39 \times 10^{-10}$, which represents  5.2\% of $\amuh$. The last
two data columns yield
$\amuh (I=0, s < 1.05~{\rm GeV})=71.53 ~ 10^{-10}$, {\it i.e.} 12.6\% of $\amuh$.
So, together, IB and I=0 effects amount to 17.8 \% of $\amuh$ in the HLS energy range.

\vspace{0.5cm}

Dealing with the photon terms is a more delicate matter and might introduce
a strong model dependence while the value for $\amuh(\gamma\&[I=1],s < 1.05~{\rm GeV})$
in  Table \ref{tab:gm2_hls} can reasonably be trusted.

Indeed, while IB is canceled out, 
 the photon coupling to a pion pair within BHLS is $ g_{\gamma \pi\pi}^{HLS}=(1-a_{HLS}/2)e$,
which numerically gives $g_{\gamma\pi\pi}^{HLS} \simeq -0.25 e$; in standard VMD models one assumes
$ g_{\gamma\pi\pi}^{HLS}=0$ ({\it i.e.} $a_{HLS}=2$) while  models based on
scalar QED (as \cite{{Jegerlehner:2011ti},Jegerlehner:2015stw}) yield good
fits with $ g_{\gamma\pi\pi}=e$.
As such kinds  of models can satisfactorily describe the pion form factor
in the $e^+e^-$ annihilation, one can legitimately suspect  that some kind of numerical
conspiracy is at work within fits when sharing physical effects between
$\gamma \ra {\rm hadr.}$ and $\rho^0 \ra {\rm hadr.}$.
With this proviso in mind, we give below  the outcome of
setting $ g_{\gamma\pi\pi}^{HLS}=0$ within our model results while reconstructing  
the amplitudes.

 \begin{table}[h]
\centering
\caption{Specific  channel contributions to the I=1 $\amuh$
The LO-HVP contribution in terms of moments in units of $10^{-10}$.
One observes that the effect of the full I=1 amplitude is almost saturated by
 the $\pi^+ \pi^-$  channel and  a very 
small correction is provided by the $\pi^0 \gamma$  and $\eta \gamma$ channels.
In the last data column, the direct coupling $\gamma \ra {\rm hadr.}$ is canceled out
together with the  final state radiation (FSR) effects .}
\label{tab:gm2_brk}
{\footnotesize
\begin{tabular}{l||c||c|c||c}
\noalign{\smallskip}\hline\noalign{\smallskip}
\phantom{moment} 
&\multicolumn{3}{c}{$\amuh(I=1 ~\& ~\gamma)$ } & $\amuh(I=1)$ \& no $\gamma$\\
\phantom{moment} &
{All HLS Channels} &
{$\pi^+ \pi^-$} &
{$\pi^+ \pi^-$ + $\pi^0 \gamma + \eta \gamma$} &
{$\pi^+ \pi^-$}  
\\
\noalign{\smallskip}\hline\noalign{\smallskip}
 $\amuh(0)$&    552.15$\pm$  0.75&  551.81$\pm$ 0.75 & 552.14 $\pm$  0.75 	&  562.45 $\pm$  0.78	\\
 $\amuh(1)$&    489.66$\pm$  0.67&  489.37$\pm$ 0.67 & 489.65 $\pm$  0.67	&  498.55 $\pm$  0.68   \\
 $\amuh(2)$&    473.49$\pm$  0.65&  473.20$\pm$ 0.64 & 473.48 $\pm$  0.65	&  481.60 $\pm$  0.67   \\
 $\amuh(3)$&    469.70$\pm$  0.64&  469.42$\pm$ 0.64 & 469.70 $\pm$  0.63	&  477.51 $\pm$	 0.66  	\\
 $\amuh(4)$&    468.62$\pm$  0.64&  468.34$\pm$ 0.64 & 468.61 $\pm$  0.64	&  476.35 $\pm$	0.66 \\
\noalign{\smallskip}\hline\noalign{\smallskip}  
  $\amuh$ &     468.03$\pm$  0.65&  467.75$\pm$  0.64& 468.02$\pm$   0.64	&   475.70 $\pm$0.66	\\
\noalign{\smallskip}\hline
\end{tabular}
}
\end{table}

In Table \ref{tab:gm2_brk} we display information aiming at substantiating the contributions
other than $\pi^+ \pi^-$ to $\amuh(I=1)$ -- including/excluding the $\gamma \ra {\rm hadr.}$ vertex 
contributions. Integrated over our energy range
 of interest, the $\pi^0 \gamma$ channel contributes  $\simeq 2~10^{-11}$,
 the $\eta \gamma$ channel $\simeq 10^{-11}$, while the $\pi^+ \pi^-\pi^0$
 and $K \overline{K}$ channels provide contributions at the $\simeq 10^{-12}$
 level or less. 
 
 The last data column in Table \ref{tab:gm2_brk}
 displays the $\rho$ term contribution only, {\it i.e.} one has canceled out the
 $\gamma \ra {\rm hadr.}$ vertex, and also -- for consistency -- the FSR contribution; 
 in this case all channels except for $\pi^+ \pi^-$ give invisible contributions to the 
 various $\amuh(n)$'s listed. It is also interesting to notice that 
  the main effect of
 the photon couplings is to reduce the values for the  $\amuh(n)$'s. 
 
 One should also remind the existence of channels missing the BHLS framework 
 \cite{Benayoun:2012wc} which contribute $(1.34 \pm 0.11) \times 10^{-10}$
 to $\amuh$. This represents a systematic error clearly of limited influence.

Finally, Table \ref{tab:mom_brk} reports on our numerical results for
$\cM(-n)$ and $\tM(-n)$; the first two data columns 
in the upper part of this Table are
a copy out of the central data columns in Table \ref{tab:mom}
and are reminded for convenience. The third data column shows
the effect of only canceling out IB effects -- as identified within BHLS. 
The  two rightmost data columns in the lower part of this Table
give resp. the moments  when keeping the I=1 part of the amplitude {\it and} the
photon terms (direct coupling + FSR) and the last one only when keeping
the I=1 component (remind the proviso expressed above). These numbers
reflect the same phenomena as commented just above for $\amuh$.

\begin{table}
\caption{Moments of the $\amuh$-expansion in units of $10^{-5}$. Here
$\cM(-n)$ and $\tM(-n)$ are evaluated via Eqs. (\ref{Mint})
and (\ref{tildeMdire}) in terms of $R(s)$ as provided by
$e^+e^-$--annihilation data and/or predictions of the BHLS model Lagrangian.
The integration lower limit is $m_{\pi^0}^2$ and the upper limit is $(1.05~\gv)^2$.}
\label{tab:mom_brk}
\hspace{-2cm}
\centering
{\scriptsize
\begin{tabular}{lr@{.}lr@{.}l|r@{.}lr@{.}l|r@{.}lr@{.}l}
\noalign{\smallskip}\hline\noalign{\smallskip}
moments &
\multicolumn{4}{c|}{Exp. Value} &
\multicolumn{8}{c}{HLS model}  
\\
HLS scope &
\multicolumn{4}{c|}{$~~~~$} &
\multicolumn{4}{c}{Standard Fit} &
\multicolumn{4}{c}{$\gamma$ \& I=0 \& I=1 (IB removed)} 
\\
\noalign{\smallskip}\hline\noalign{\smallskip}
$\cM(0)  $
& 8&6275&$\!\!\!\!\!\pm$0&0495
& 8&6041 &$\!\!\!\!\!\pm$0&0130
& 8&1613&$\!\!\!\!\!\pm$0&0122
\\
$\cM(-1) $
& 0&22944&$\!\!\!\!\!\pm$0&00184
& 0&23197&$\!\!\!\!\!\pm$0&00031
& 0&22023&$\!\!\!\!\!\pm$0&00029

\\
$\cM(-2) $
& 0&008669&$\!\!\!\!\!\pm$0&000115
& 0&008974&$\!\!\!\!\!\pm$0&000011
& 0&008542&$\!\!\!\!\!\pm$0&000010
\\
$\cM(-3) $
& 0&0004850&$\!\!\!\!\!\pm$0&0000093
& 0&0005147&$\!\!\!\!\!\pm$0&00000064
& 0&0004902&$\!\!\!\!\!\pm$0&0000006
\\
$\cM(-4) $
& 0&00003676&$\!\!\!\!\!\pm$0&00000083
& 0&00003956&$\!\!\!\!\!\pm$0&00000005
& 0&0000376&$\!\!\!\!\!\pm$0&000000040
\\
\hline\noalign{\smallskip}
$\tilde{\cM}(-1)$
& -0&79611&$\!\!\!\!\!\pm$0&00501
& -0&80054&$\!\!\!\!\!\pm$0&00113
& -0&75948&$\!\!\!\!\!\pm$0&00103
\\
$\tilde{\cM}(-2)$
& -0&026644&$\!\!\!\!\!\pm$0&000294
& -0&027334&$\!\!\!\!\!\pm$0&000035
& -0&026009&$\!\!\!\!\!\pm$0&000032
\\
$\tilde{\cM}(-3)$
& -0&0013149&$\!\!\!\!\!\pm$0&0000228
& -0&0013847&$\!\!\!\!\!\pm$0&0000017
& -0&0013193&$\!\!\!\!\!\pm$0&0000015
\\
$\tilde{\cM}(-4)$
& -0&00009063&$\!\!\!\!\!\pm$0&00000199
& -0&00009725&$\!\!\!\!\!\pm$0&00000012
& -0&00009253&$\!\!\!\!\!\pm$0&00000010
\\
\noalign{\smallskip}\hline
\noalign{\smallskip}\hline\noalign{\smallskip}
moments &
\multicolumn{4}{c|}{Exp. Value} &
\multicolumn{8}{c}{Standard HLS Fit (Breaking effects removed)}  
\\
HLS scope &
\multicolumn{4}{c|}{$~~~~$} &
\multicolumn{4}{c}{ $\gamma$ \& I=1} & 
\multicolumn{4}{c}{I=1 \& $ \not \! \gamma$} \\
\noalign{\smallskip}\hline\noalign{\smallskip}
$\cM(0)  $
& 8&6275&$\!\!\!\!\!\pm$0&0495
& 7&1313&$\!\!\!\!\!\pm$0&0096
& 7&2635&$\!\!\!\!\!\pm$0&0100
\\
$\cM(-1) $
& 0&22944&$\!\!\!\!\!\pm$0&00184
& 0&20514&$\!\!\!\!\!\pm$0&00027
& 0&21368&$\!\!\!\!\!\pm$0&00030

\\
$\cM(-2) $
& 0&008669&$\!\!\!\!\!\pm$0&000115
& 0&008297&$\!\!\!\!\!\pm$0&000010
& 0&008857&$\!\!\!\!\!\pm$0&000012
\\
$\cM(-3) $
& 0&0004850&$\!\!\!\!\!\pm$0&0000093
& 0&0004839&$\!\!\!\!\!\pm$0&0000005
& 0&0005252&$\!\!\!\!\!\pm$0&0000008
\\
$\cM(-4) $
& 0&00003676&$\!\!\!\!\!\pm$0&00000083
& 0&00003690&$\!\!\!\!\!\pm $0&000000040
& 0&00004024&$\!\!\!\!\!\pm$0&000000058
\\
\hline\noalign{\smallskip}
$\tilde{\cM}(-1)$
& -0&79611&$\!\!\!\!\!\pm$0&00501
& -0&69637&$\!\!\!\!\!\pm$0&00093
& -0&72000&$\!\!\!\!\!\pm$0&00100
\\
$\tilde{\cM}(-2)$
& -0&026644&$\!\!\!\!\!\pm$0&000294
& -0&025023&$\!\!\!\!\!\pm$0&000031
& -0&026517&$\!\!\!\!\!\pm$0&000036
\\
$\tilde{\cM}(-3)$
& -0&0013149&$\!\!\!\!\!\pm$0&0000228
& -0&0013001&$\!\!\!\!\!\pm$0&0000015
& -0&0014056&$\!\!\!\!\!\pm$0&0000020
\\
$\tilde{\cM}(-4)$
& -0&00009063&$\!\!\!\!\!\pm$0&00000199
& -0&00009149&$\!\!\!\!\!\pm$0&00000010
& -0&00009980&$\!\!\!\!\!\pm$0&00000014
\\
\noalign{\smallskip}\hline
\end{tabular}
}
\end{table}

\section{Conclusion}
\label{sec:conclusion}
We have demonstrated that the Mellin-Barnes moments expansion for
$\amuh$ works surprisingly well (see Fig.~\ref{fig:gm2moments}),
exhibiting a fast convergence with 4 or 5 moments only. In the
timelike approach it assumes  the non-perturbative $R(s)$ 
given for low $s$ and in resonance regions, while the known well-behaved
integral kernel $\hat{K}(s)$ of (\ref{lohadalt}) is
expanded. Obviously, when $R(s)$ is given the moment expansion is just
more elaborate than simply calculating the integral (\ref{lohadalt})
directly. However, in lattice QCD calculations, which are constrained
to the Euclidean (spacelike) region, rather than $R(s)$ which is far
from being accessible there, the primary object is
the electromagnetic current correlator in
configuration space 
\be
\langle J^\mu(\vec{x},t)\,J^\nu(\vec{0},0)\rangle\,,
\label{ECC}
\ee 
where \mbo{J^\mu(\vec{x},t)} is the electromagnetic current, 
and various types of moments are extractable from it.
Concerning $\amuh$ the integral representations (\ref{ADI}) and (\ref{RAI})
reveal that the Euclidean vacuum polarization function $\Pi(Q^2)$ in momentum space
is what is needed. So, in principle, a Fourier transform like 
\ba
\Pi(Q^2)\,\left(Q^\mu Q^\nu- \delta^{\mu\nu} \,Q^2\right)=\int
\D t\, \E^{\omega t}\,\int \D^3 \vec{x}\,
\E^{\I\,\vec{q}\,\vec{x}}\,\langle J^\mu (\vec{x},t)\,J^\nu(\vec{0},0)\rangle
\label{HVPanalytic}
\ea
\mbo{Q=(\vec{q},-\I\,\omega)} \mbo{\vec{q}} is a spatial momentum and \mbo{\omega} 
the photon energy (input), provides the object required. A Fourier
transformation of lattice data, however, is far from being easy and
uncertainties due to fluctuations in general turn out to be large (see
e.g.in~\cite{Jansen:2014nta}). Moment expansions therefore are often a
way out for getting more precise estimates of the HVP function.  The
Taylor expansion approach (\ref{Taylor}) with coefficients given by
(\ref{latticeTayCoe}) (see
also~\cite{Francis:2014dta,Golterman:2014rda,Gregory:2015bno}) in
conjunction with Pad\'e approximants used
in~\cite{Chakraborty:2016mwy}, is the simplest one can do. However,
the low order Pad\'e approximants we get with 4 moments, illustrated
in Fig.~\ref{fig:HVPPades}, are not very convincing for larger momenta
in a region which still gives a non-negligible contribution to the
$\amuh$ integral (\ref{RAI}). 

The Mellin-Barnes moment approach is
more promising but also much more elaborate. The reason is that
besides the lattice QCD accessible Taylor moments $\cM(-n)$ we also
need the moments $\tM(-n)$. The latter, in the Euclidean regime,
require in addition to extract the lattice QCD accessible moments $\Sigma(-n;
s_0)$. To our knowledge such an analysis has not yet been
performed so far by lattice QCD groups. Our analysis shows that a reliable
extraction of the log suppressed $\tM(-n)$ is difficult, the main
problem being the need for Pad\'e approximants to extend the low
energy expansion towards higher energies. Unfortunately, Pad\'e approximants (PA)
cannot match QCD asymptotics, which means that one has to use an
appropriate cutoff where one can continue the PA with pQCD predictions.
In the Euclidean region such a cutoff is expected to be around 
2.5~GeV, as one observes by confronting the data extracted Adler function
with its pQCD prediction~\cite{EJKV98,Jegerlehner:2008rs}. It means
that it is advised to cut the PA of Fig.~\ref{fig:HVPPades} at 2~GeV
to 2.5~GeV in any case and use $\dalh(-Q^2)$ for $Q>Q_1$ or its pQCD
prediction. This also has the advantage that one may use the Pad\'e pairs
[n,n] and [n,n+1] as upper and lower bounds, keeping integrals convergent. 

Note that in order to determine the two sets of moments $\cM(-n)$ and
the $\tM(-n)$, in both cases, in Minkowski space, where we work with
data, and in Euclidean space, where we work with Euclidean current
correlators, we need and have available only one quantity either
$R(s)$ in the first case or $\Pi(Q^2)$ in the second case. The problem
on the lattice is that the ``trailhead'' is always the Euclidean
configuration space correlators (\ref{ECC}), which allows us to get
directly the Taylor moments $\cM(-n)$ via
(\ref{latticeTayCoe}). However, a corresponding direct configuration
space evaluation of the moments $\Sigma(-n; s_0)$, needed to obtain
the moments $\tM(-n)$, seems not to exist, which means that we have to
get $\Pi(Q^2)$ in any case first, by Fourier transformation of the
primary configuration space correlators. 

Fortunately, it turns out that our master formula
(\ref{tildeMlattice}) for extracting $\tM(-n)$ allows for a stable and
accurate estimation of the log suppressed moments. This is illustrated
in Fig.~\ref{fig:tildeMofs0}. As an input, an extended (to higher
orders) set of Taylor moments $\cM(-n)$ is sufficient to allow us to
construct the required truncated HVP function to which we have to
apply the Pad\'e improvement via [n-1,n]+[n,n] pairs up to some -- not too
high -- cutoff, above which one can include the high energy tail as
predicted by pQCD.

Although the moments expansions seem to work surprisingly well, we
have some reservation concerning calculating $\amuh$ in terms of
moments.  The moments method emphasizes the low momentum region below
the $\rho$ resonance with $m_\mu$ as a reference scale. In the
standard representations (\ref{lohadalt}) the low energy region also
gets enhanced by $1/s^2$, but not more, and we know the $\rho$
resonance yields the dominant piece. It thus properly weights the
$M_\rho$ mass region but also has the right high energy behavior to
get the integral converge. Our concern is that neither in evaluations
based on data nor in lattice QCD estimates (see Figure 5 of
\cite{Boyle:2011hu} and Figure 1 of \cite{Aubin:2015rzx}
for recent discussions of that point\footnote{Present simulations
reach typically $Q_{\rm min}=2\pi/L$ with $m_\pi L\gapprox 4$ for
$m_\pi\sim 200~\mv$, such that $Q_{\rm min}\sim 314~\mv$ while the
kernel displayed in Fig.~\ref{fig:ADIkernels} shows the peak at about
$150~\mv$.}) the low energy tail is easy to get very precisely and
therefore the moments expansion tends to increase uncertainties by
giving high weight to the problematic region.  The problem in lattice
QCD of course is that $R(s)$ is not available such that a priori the
Euclidean representations (\ref{RAI}) or (\ref{ADI}) come into
play. In the Euclidean region there are no resonances and no flavor
thresholds and the structures characterizing the timelike region
appear completely smoothed out by the dispersion integrals
(\ref{HVPDR}) or (\ref{DI}). So we are confronted with the question
where the dominant contributions come from in these representations.
The answers are given by Figs.~\ref{fig:RAIkernels} and
\ref{fig:ADIkernels} which show pronounced peaks in the distributions
below the 1~GeV scale. To be more precise, the contribution to $\amuh$,
in the $x$-integral representations (\ref{ADI}) or (\ref{RAI}), is the area under the curve
shown in the left panels of Figs.~\ref{fig:ADIkernels} and~\ref{fig:RAIkernels},
respectively. The figures illustrate the role of extrapolations
(especially the large volume limit) still required in order to obtain the
bulk of $\amuh$. What is used is of course shape information from chiral
perturbation theory and from vector meson dominance model type
parametrizations which help to control the extrapolation fairly well.

At the end the key problem is how to extract from lattice data of the
Euclidean configuration space current correlator (\ref{ECC}) a
reliable Euclidean HVP function $\Pi(Q^2)=\Pi_{\rm bare}(Q^2)-\Pi_{\rm bare}(0)$
or, better,  the Adler function $D(Q^2)$, the latter being devoid of
problems related to UV subtraction term $\Pi_{\rm bare}(0)$ and which is bounded
in the high energy limit\footnote{Note that on a lattice in a finite volume
$\Pi_{\rm bare}(0)$ is a difficult object as it is a zero momentum object
depending on the lattice spacing and thus requires careful
extrapolations $L\to\infty$ and $a \to 0$. $L$ the box extension and
$a$ the lattice spacing.}. The Taylor + Pad\'e approximants (TPA) method is
more sensitive to the high energy tail as becomes obvious from
Fig.~\ref{fig:HVPPades} in conjunction with Fig.~\ref{fig:gm2moments}.
The corresponding TPA procedure applied to the Adler function to $Q^2$
ratio $D(Q^2)/Q^2$, which has a finite limit $Q^2\to 0$ and behaves as
$1/Q^2$ at higher energies, for the appropriate Pad\'es definitely
exhibits a much better behavior in this respect (see
Fig.~\ref{fig:ADPPades}) and allows to reduce the uncertainty accordingly.

The Mellin-Barnes moments method is
much more elaborate since, besides the Taylor coefficients $\cM(-n)$,
the log weighted moments $\tM(-n)$ are required, which in lattice QCD are much more
difficult to evaluate. A major difference between the TPA and the MBM
methods seems to be that the MBM method consists in the
expansion of the known integral weight function, not touching the real
object of concern, the non-perturbative object $R(s)$. The trick is to
focus on $R(s)$ by reweighting it with a series of different magnifying
filters. In contrast, the TPA method is based on a low momentum
expansion of the non-perturbative object, the Euclidean $\Pi(Q^2)$ or $D(Q^2)/Q^2$
itself.

We think that the use of Pad\'e approximants is not optimal in our context,
because the proper QCD high energy behavior cannot be obtained by
Pad\'eization of the non-perturbative low energy tail. The upper bounds
[n,n] lead to an UV singularity such that only lower bounds [n-1,n]
actually can be accepted. To approach the solution one has to
consider the convergence of the series [n-1,n] for n=2,3,4,... 
Pad\'e approximants can be very useful to bridge (interpolate) between a low
energy and a high energy expansion, as it works very well for the massive
3-loop Adler function (see ~\cite{EJKV98} for details), for example.

We therefore advocate to use the integral representations, preferably
(\ref{ADI}), directly, as
e.g. in~\cite{DellaMorte:2014rta,DellaMorte:2016izp} and to determine
the vacuum polarization function and/or the Adler function as
precisely as possible. For both objects $\Pi(Q^2)$ and $D(Q^2)$ rather
precise reference functions are available obtained by standard
analysis of $R(s)$ data in conjunction with pQCD. Nevertheless, checks
with the help of moment expansions are useful to make sure that the
obtained vacuum polarization functions are under control.  The present
analysis provides the ``data for the moments'' to perform such
crosschecks. Needless to say that the moment analysis is much more
elaborate than performing the basic integrals once directly. Since the
low momentum region is difficult to evaluate in lattice QCD, the
minimum momentum on the lattice is $2\pi/L$ where $L$ is the lattice
box length. So the access of low momenta is via extrapolation to the
infinite volume limit. A promising possibility is the method of analytic
continuation proposed in~\cite{Jansen:2014nta}, which allows to access
low momenta by interpolation, rather than be extrapolation. In this
approach one computes the HVP function (\ref{HVPanalytic}) and varies
\mbo{\omega} as an input parameter and obtains a smooth function for 
\mbo{\Pi(Q^2=-\omega^2+\vec{q}^{\,2})}. Spacelike and timelike momentum
regions can be covered and one can reach small momenta and even zero momentum.
This is supposed to work under the condition that
\ba
-Q^2=\omega^2-\vec{q}^{\,2} < M_V^2\;,\;\; \mathrm{ \ or \ \ } \omega <
M_V \,,
\ea
where $V$ is the lowest vector state, the lattice realization of the
$\rho$ meson in a given simulation. 

Unfortunately present computer resources do not yet admit to get
precise results in the extended range of interest because simulation
data are still too noisy~\cite{Jansen:2014nta}, but for the future the
method looks very promising.
 
 \vspace{0.5cm}
 
It should be noted that so far lattice evaluations of the LO $\amuh$
are mainly based on conserved isovector current calculations and do not include iso-singlet
effects (see ~\cite{Francis:2014hoa,Blum:2015you} and references therein, however), 
isospin breaking effects like $\rho-\omega$ mixing, and
electromagnetic effects like $\rho-\gamma$ mixing~\cite{Jegerlehner:2011ti}
or hadronic final state radiation. All these effects are incorporated
in our data and in the corresponding BHLS model and fits as described 
 in Sect.~\ref{sec:BHLSisovec}, if not stated otherwise.

Another question concerns the possible model dependence of the results
obtained with the BHLS effective field theory. As the global fit
quality of the NSK+KLOE+BESIII data is surprisingly good, it is
unlikely that a different or improved model would be able to improve
the global fit quality substantially. Actually, different
implementations of the Resonance Lagrangian Approach (RLA) are
expected to be equivalent provided the high energy behavior is
adjusted to be consistent with
QCD~\cite{Ecker:1988te,Ecker:1989yg}. Without actually performing a
corresponding analy\-sis, e.g. by including additional higher order
corrections or by using a different implementation of a RLA model,
adding some error would be a plain guess.

In our opinion a model error is already included in our fit errors,
since if the model is mismatching with parts of the data of course
this is reflected in the global fit error. As an example we mention
that dropping photonic corrections from our BHLS model, or not
including photonic corrections in any of the alternative RLA
implementations, would spoil the good agreement between $\tau \to
\pi^\pm \pi^0 \nu_\tau$ spectral data supplemented by the isospin
breaking effects on the one hand and the $\epm \to \pi^+\pi^-$ data on
the other hand, as documented by Fig.~\ref{taupred}. In other words,
neglecting relevant photon-hadron couplings and corresponding loop
effects (in self-energies at least), which implies substantial
$\gamma-\rho^0$ mixing effects among others, would ruin the excellent
BHLS global fit quality. This also implies that photonic corrections
have to be included in LQCD calculations at some stage. Taking into
account isospin breaking effects originating from the difference in
the $u$ and $d$ quark masses, like $\omega-\rho$ mixing, as well as
electromagnetic effects, is in progress, but is by far not a simple task.


\section*{Acknowledgments}
 F.J. thanks the Laboratori Nationale di Frascati (INFN - LNF) for the
 kind hospitality extended to him. F.J. also thanks Rainer Sommer,  
 Gregorio Herdo\'iza and  Maria Paola Lombardo for helpful discussions.

\end{document}